\documentclass[twocolumn,final]{elsarticle}
\usepackage{multirow,array,hhline}
\usepackage{framed,multirow}

\usepackage{graphicx}
\usepackage{geometry}
\usepackage{babel}
\usepackage{amsmath}

\usepackage{subcaption}
\usepackage{amssymb}
\usepackage{latexsym}
\usepackage{url}
\usepackage{xargs}
\usepackage{hyperref}

\providecommand{\tabularnewline}{\\}

\begin{document}

\title{A vascular synthetic model for improved aneurysm segmentation and detection via Deep Neural Networks}

\author[1]{Rafic {Nader}$^\dagger$}
\author[1,2]{Florent {Autrusseau}\corref{cor1}$^\dagger$}
\ead{Florent.Autrusseau@univ-nantes.fr}
\cortext[cor1]{Corresponding author: 
 Tel.: +33-240683156; $^\dagger$ Authors RN and FA contributed equally.}
\author[1,3]{Vincent {L'Allinec}}
\author[1]{Romain {Bourcier}}

\address[1]{Nantes Université, CHU Nantes, CNRS, INSERM, l’institut du thorax, F-44000 Nantes, France}
\address[2]{Nantes Université, Polytech’Nantes, LTeN, U-6607, Rue Ch. Pauc, 44306, Nantes, FRANCE}
\address[3]{CHU Angers, Service Radiologie, 4, rue Larrey, 49933, Angers, FRANCE}

\begin{abstract}
We hereby present a full synthetic model, able to mimic the various constituents of the cerebral vascular tree: the cerebral arteries, the bifurcations and the intracranial aneurysms. 
By building this model, our goal was to provide a substantial dataset of brain arteries which could be used by a 3D Convolutional Neural Network (CNN) to either segment or detect/recognize various vascular diseases (such as artery dissection/thrombosis) or even some portions of the cerebral vasculature, such as the bifurcations or aneurysms. 
In this study, we will particularly focus on Intra-Cranial Aneurysm (ICA) detection and segmentation. 
The cerebral aneurysms most often occur on a particular structure of the vascular tree named the Circle of Willis. Various studies have been conducted to detect and monitor the ICAs and those based on Deep Learning (DL) achieve the best performances. Specifically, in this work, we propose a full synthetic 3D model able to mimic the brain vasculature as acquired by Magnetic Resonance Angiography (MRA), and more particularly the Time Of Flight (TOF) principle. Among the various MRI modalities, the MRA-TOF allows to have a relatively good rendering of the blood vessels and is non-invasive (no contrast liquid injection). 
Our model has been designed to simultaneously mimic the arteries geometry, the ICA shape and the background noise. The geometry of the vascular tree is modeled thanks to an interpolation with 3D Spline functions, and the statistical properties of the background MRI noise is collected from MRA acquisitions and reproduced within the model. In this work, we thoroughly describe the synthetic vasculature model, we build up a neural network designed for ICA segmentation and detection, and finally, we carry out an in-depth evaluation of the performance gap gained thanks to the synthetic model data augmentation.
\end{abstract}


\maketitle

\section{Introduction}

This work has been carried out in the context of a wide medical research project in which neuroradiologists intend to estimate the risk of occurrence and/or rupture of Intra-Cranial Aneurysms~\citep{Bourcier2017,Neurosurg2020}.
The advent of ICA formation results from various factors, among which the genetic risk seems predominant \citep{Zhou2018,Bourcier2018}. 
However, it is commonly accepted among physicians that the geometric disposition of the cerebral vascular tree might explain why a weakened vessel wall (due to genetic or environmental factors) might give rise to an aneurysm. 

Untreated brain aneurysms pose a significant risk of rupture, which can result in a hemorrhagic stroke. In fact, this rupture can potentially lead to the patient's death in as much as $50\%$ of all cases. 
The MRA-TOF modality  is frequently used for aneurysms detection \citep{sailer2014}. Unlike other methods like Digital Subtraction Angiography (DSA) and Computed Tomographic Angiography (CTA), MRA-TOF is radiation-free and doesn't require the administration of a contrast agent~\citep{Summerlin2022,Adams99}. Given the mounting workload and the demanding nature of the detection process undertaken by radiologists, there is an increasing  need for an automated tool to detect and monitor aneurysms at an early stage. 
Prior to the widespread adoption of Deep Learning (DL), research studies employed imaging filters or traditional machine learning techniques to detect aneurysms \citep{LAURIC2010149,zeng,yang2011computer,arimura2006computerized,nemoto2017feasibility,hanaoka2019}. 
Recent advances in artificial intelligence, particularly those involving deep Convolutional Neural Networks (CNNs), have significantly enhanced the development of automatic tools in the field of medical imaging \citep{wang2012machine}.
To date, several deep learning based approaches have been proposed for ICA segmentation and/or detection \citep{stember2019,Park2019,chen2020,faron2020,joo2020,Timmins2021,natu,yang2021deep,IVANTSITS2022102333}. Among these, the ADAM Challenge, hosted in conjunction with MICCAI 2020, compared 11 different DL approaches for detecting or/and segmenting aneurysms on MRA-TOFs. 
The algorithm \citep{baumgartner2021} that secured the top position achieved notable results, with an average sensitivity of $0.67$ and a false positive rate of $0.13$. 
It is important to note that a majority of the existing methods have been formulated using private clinical data that comes with meticulously refined manual annotations. Indeed, one of the obstacles in developing deep learning methods for medical imaging applications is the lack of large annotated datasets, particularly for the segmentation task. To mitigate this, Di Noto et al. \citep{di2023} proposed  the use of ``weak'' annotations and they obtained good results with an average lesion sensitivity of $0.83$ and a false positive rate of $0.8$.

The rationale behind our work is to try to reduce as much as can be, or even possibly to free oneself from any manual labeling. In other words, we expect that using several hundreds or even thousands of modeled bifurcation to train a network might provide better performances than using only  a couple of hundreds actual MRA-TOF segmentations.
Unlike previous works, in our approach, we investigate the brain aneurysm detection task by exploiting synthetic data. 
While data augmentation stands out as a well-known technique for augmenting the number of data samples, its application requires careful consideration. 
In the context of medical images, such image manipulations might tamper with the geometrical or statistical properties in an undesirable way, \textit{i.e.} render the augmented images too distant from their corresponding ground truth. 

Indeed, for a dataset augmentation, the first thought that naturally comes to mind is to simply use classical 3D geometrical deformations (3D grid tweaking). However, among the pitfalls encountered when using classical geometrical distortions onto the 3D patches to train a 3D U-Net for vascular segmentation, we have witnessed a deformation of the arteries, which no longer resembled to actual cerebral vessels as the elastic deformations could lead to a flattening or exaggerate stretching of certain portions of the 3D tubes. Moreover, a geometric deformation applied onto a 3D MRA-TOF patch will inevitably distort the background noise in an undesirable way, thus changing the noise distribution and its statistical features.
Hence, applying classical elastic deformations \citep{Ronneberger2015,cicek20163d} might actually be counterproductive and reduce the CNN performances.

In the past, about two decades ago, several works have been devoted to the design of computer models intended to mimic arterial trees. At that time, the studies focused on constrained constructive optimization~\citep{Karch1999}. Some models were particularly designed to offer a high graphical fidelity through a better understanding of the biophysical properties~\citep{Szekely2002}. A relatively nice rendering was obtained on liver vascular trees for instance in~\citep{Coatrieux2003}.
Such models were mostly designed in the aim to study angiogenesis (physiological process leading to the formation of new blood vessels.).
More recently, the VascuSynth model~\citep{Hamarneh2010} was proposed in the aim to produce vast amounts of volumetric vasculature images. Here, the aim was different, the authors intended to generate a synthetic  dataset for image segmentation. Indeed the modeled images were generated along with their underlying ground truth segmentation. 
All these computer models achieved a quite accurate modeling of the acquired medical images (mostly trying to mimic liver or lung vasculatures). 
However, modeling the cerebral vascular tree might be a bit more challenging, as the arteries are commonly longer, and may exhibit a stronger tortuosity. 
Moreover, in our study, the goal strongly differs. We intend to generate vast amounts of images to train a neural network for a pattern recognition task.

In previous works \citep{EMBC2021}, we have proposed a method aiming to ``humanize'' some mice $\mu$-CT acquisitions. In this work, some 2D image patches were collected onto the micro-scans of mice brains and noise was added in the background to make the mice arteries resemble as much as possible to human arteries as acquired via MRA-TOF acquisitions. These ``humanized'' acquisitions were utlimately used to train a 2D CNN for vascular segmentation.

More recently, we have devoted a subsequent work \citep{ICPR2022} to a new approach, where we no longer intended to exploit an inter-species framework; our aim in this study was to build a fully synthetic model of 3D bifurcations and Intra-Cranial Aneurysms. Here, when creating the geometrical shape of the bifurcation, we followed a 5 steps process: \textit{i)} Four different 3D coordinates points were selected onto a $60\times60\times60$ 3D patch (the bifurcation center, and its three extremities), \textit{ii)} linear segments were drawn to link those 4 points (forming a bifurcation), \textit{iii)} a convolution with a 3D spherical kernel was performed on each and every branch, the sphere radius could evidently vary in order to model the various branches' diameters, \textit{iv)} a geometrical distortion (elastic deformation \citep{Ronneberger2015,Simard2003}) was applied onto the so-obtained thickened bifurcation, and finally, \textit{v)} a background noise was constituted and added around the so-obtained 3D bifurcation model. Moreover, in this work, we also proposed an Intra-Cranial Aneurysm model that could be superimposed onto the bifurcation. The ICA could also be distorted (via the same elastic deformation method), and annexed onto the bifurcation model. A particular attention was devoted to the generation of a plausible background noise via the use of Gaussian Mixture Models (GMM).

Although this initial model \citep{ICPR2022} proved to be able to mimic 3D bifurcations, it showed some limitations, and notably when it comes to recognize a given bifurcations of interest within the Circle of Willis (CoW). Moreover, the aneurysm location tended to present some slight shifts away from the bifurcation wall. Such small displacements might be sufficient to slightly lessen the ICA detection performances.

In the current work, our model accuracy is considerably increased, and we intend to propose a set of much more realistic bifurcations and aneurysms.
Since our intended application involves identifying an aneurysm within a specific bifurcation or artery from an MRA-TOF scan, it is crucial to accurately model various essential arterial features: the shape, orientation, diameters, and tortuosity. As for the aneurysms, the model should allow to adjoin aneurysms of various shapes and sizes onto different locations of the bifurcation. Finally, the performances in terms of image segmentation might depend on the accuracy of the modeled surrounding background noise, hence, it is important for the model to faithfully duplicate the background noise.

This paper is organized as follows: In section~\ref{sec:MnM}, we thoroughly describe the synthetic vasculature model. Its three main features are presented, namely \textit{i)} the arteries geometry, \textit{ii)} the surrounding MRA-TOF noise and \textit{iii)} the modeled aneurysm. 
Next, in section~\ref{sec:ExpRes} we provide an in-depth description of the generated dataset, we describe the CNN architecture, evaluate both the ICA segmentation and detection performances. We try to assess the performance gain brought by using the synthetic images alongside the manually labeled ones. To do so, we run two separate experiments involving either the manually segmented images only or adjoining the modeled patches. 
Finally, in section~\ref{sec:Discussion}, we discuss the benefit of using the synthetic model for intracranial aneurysm detection and conclude this work. 

\section{Material and Method}
\label{sec:MnM}

Unlike the previous model, proposed in \citep{ICPR2022}, here we intend to come up with a full synthetic model of 3D cropped MRA-TOF portions, including not only the one bifurcation of interest, but also its whole neighborhood. 
In the framework of a bifurcation recognition task, a given target bifurcation could be recognized not only by its shape or orientation, but also by the neighboring information (adjacent arteries for instance). Such a model would be rather difficult to derive from previous existing models \citep{Hamarneh2010,EMBC2021,ICPR2022}. 
Hence, we propose a completely new method here. Within a 3D cropped portion of a MRI volume, we collect the 3D coordinates of the arteries' skeleton (centerline of the 3D tubes), and we further fit those centerlines using 3D spline functions. 

It is quite vastly admitted that there can be a significant structural variability in the vasculature of individuals \citep{Bogunovic2013,Robben2016}. Indeed, one can find relatively strong variations among the shape of the CoW (some arteries and bifurcations may be missing for some patients, or even some extra bifurcations could be present for others), and the anatomical properties of the cerebral arteries can also strongly differ from one person to another. The bifurcations' angles, the arteries' tortuosity, their diameters or even their geodesic length can significantly differ. 

\begin{figure}[!ht]
\begin{centering}
\includegraphics[width=0.6\columnwidth]{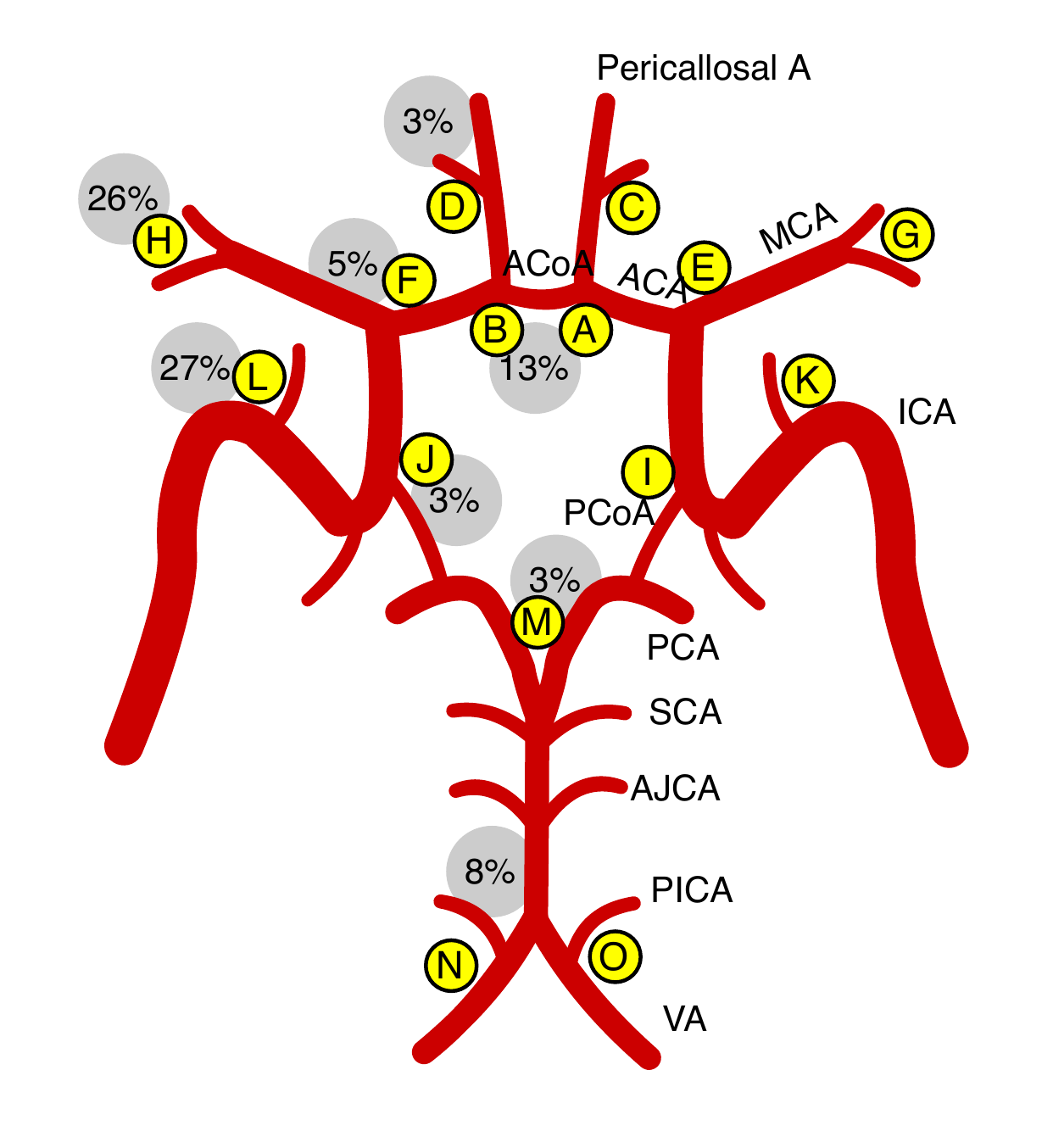}
\par\end{centering}
\caption{Schematic representation of the Circle of Willis on a human vasculature.\label{fig:SchematicWillis}}
\end{figure}

A schematic representation of the Circle of Willis is given in Fig. \ref{fig:SchematicWillis}. The letters within the yellow discs (from A to O) depict the particular bifurcations we are interested in for this study. The percentages within the gray discs represent the risk percentage of aneurysm formation.

\begin{figure*}[!ht]
\begin{centering}
\includegraphics[width=1\textwidth]{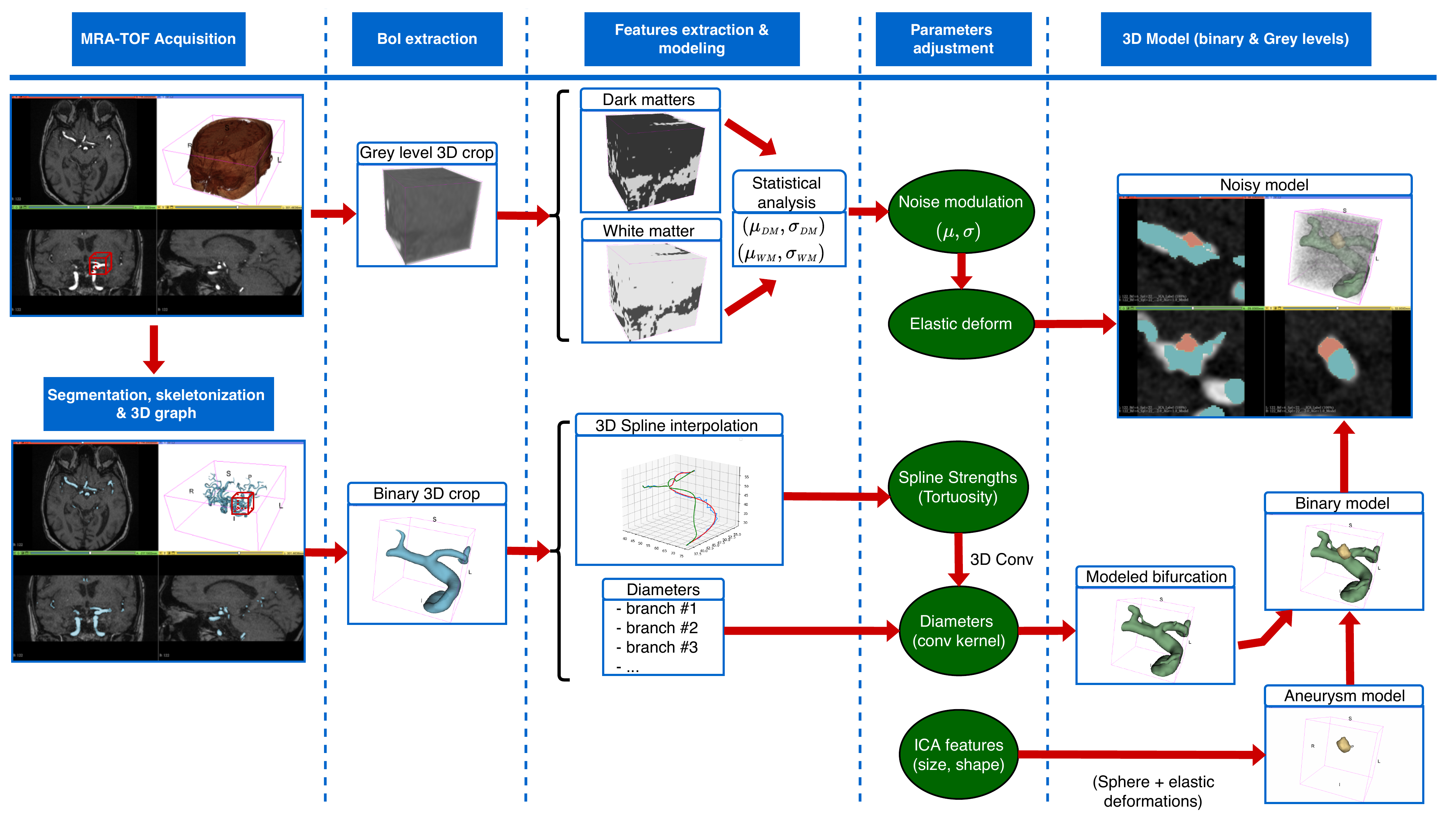}
\par\end{centering}
\caption{Schematic representation of the whole bifurcation model\label{fig:VaMosProcess}}

\end{figure*}

Such an important variability can make the task quite difficult for Neural Networks to properly recognize and/or segment the bifurcations/ arteries of interest.
That is why our aim here is to produce a model being able to generate a vast amount of relatively similar arteries, and be in a position to adjust at will, the geometric features and the grey level amplitude of the vasculature, or even the statistical properties of the surrounding noise. 
Similar to our previous project \citep{EMBC2021}, not only does our model consider the geometry of the blood vessels, but, it also provides an accurate background noise replication. 
However, both aspects are much more advanced in this new approach.
The full process of the vasculature model is represented in Fig. \ref{fig:VaMosProcess}. The blue rectangles represent the different steps needed to produce one bifurcation model, the green ellipses show the various parameters we can modulate to distort the bifurcation or its background. 

Let us first focus on the shaping of the geometrics of the bifurcation. The goal being to come up with similar shapes that can be slightly modified, but with the strong requirement to still best represent actual human arteries, the mimicking of true MRA-TOF components is crucial. 

\subsection{Modeling the arteries' geometry}
\label{ssec:ModGeomArt}

As a very first step, we use our previous works \citep{Bifurc_patent_2018,Nouri2020} to extract a 3D graph from the segmented MRA-TOF volume. 
Such a graph is simply composed of 3D curves (branches) and nodes. A node is a representation of all extremities of a branch, may they be connected to other branches (a bifurcation) or not (an end point of a given branch). From a binary segmented vasculature, a 3D graph is thus collected, this latter allows to locate a given bifurcation among the ones worthy of interest (along the CoW) and to extract a whole 3D crop around the 3D coordinates of the bifurcation's node. 
The bifurcations of interest are automatically located thanks to the works from~\citep{NADER2023102919}. 
Within this cropped portion of the segmented MRA-TOF, the set of all coordinates along the branches are collected and curve fitting is used to represent the points using 3D splines functions \citep{Splines82,Splines93}. 

Spline functions can be represented by three different characteristics:
\textit{i)} the knot-points, defining the intervals of the chunks on which the polynomials are defined, \textit{ii)} the B-Splines' (or polynomials') coefficients, and \textit{iii)} the order of the spline, (\textit{i.e.} the degree to which the fit was performed). Once these parameters have been collected for each 3D branch composing the bifurcation, it is thus relatively easy to slightly alter them in order to distort the position of the centerline coordinates. Specifically, in this model, we only alter the polynomials coefficients. 

Once the vessels centerlines have been tweaked via the spline function alteration, we shall collect the diameters of all arteries being accounted for within the 3D crop. 
This can easily be handled by using our previous vascular tree characterization tool \citep{Nouri2020}. 
Each centerline (morphological skeleton) being first tweaked by the spline alteration, can thus go through a convolution with a spherical kernel which size is adapted to the corresponding observed diameter. We then thicken each artery according to its anatomical property. Not only does such an approach allow to approximately control the artery's shape, but also to regulate its thickness and maintain a good balance between the various branches of a given bifurcation. Typically, as can be observed on Fig. \ref{fig:SchematicWillis}, the (mirror) bifurcations labeled as 'K' or 'L' are composed of two rather thick arteries and one much thinner pointing upward, whereas bifurcations 'C' or 'D' are composed of three branches exhibiting approximately the same diameters. 
As we will see within the experimental results section, our model is able model bifurcations exhibiting unbalanced thicknesses.

\begin{figure}[!ht]
\centering 
\subfloat[Spline model with weak weights]{ 
\includegraphics[width=0.45\columnwidth]{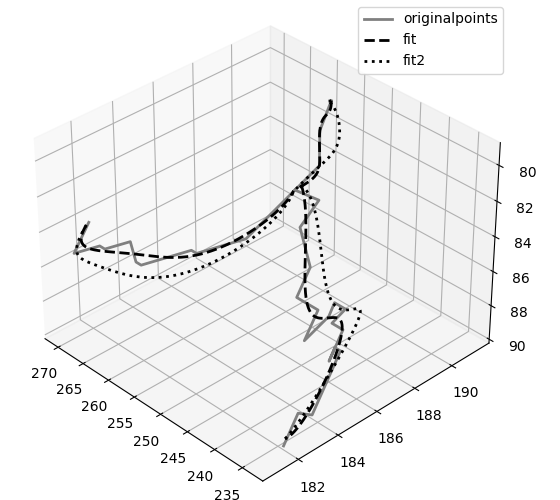}}
\hfill{}\subfloat[Spline model with larger modifications]{ 
\includegraphics[width=0.45\columnwidth]{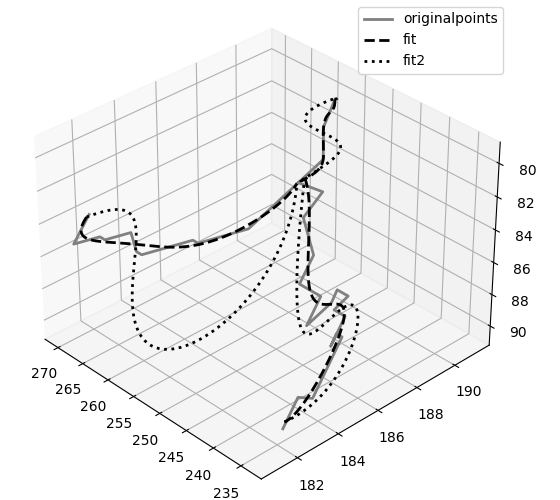}}
\caption{Examples of modified bifurcation centerlines. The solid lines represent the actual coordinates of the bifurcations' branches, the dashed lines stand for the best spline fit functions, and the dotted lines show the effect of modified spline coefficients.\label{fig:Spline_3D_bifurc}}
\end{figure}

We show on Fig. \ref{fig:Spline_3D_bifurc} some examples of 3D Spline models. The plots show, for a given bifurcation, three different 3D representation. The solid gray lines represent the actual bifurcation coordinates, as collected within the MRA-TOF acquisition, the black dashed lines stand for the spline functions that best fit the arteries, and finally, the black dotted curves show the altered spline function (the new centerline of the bifurcation to be). 

It is important to note here that, although the modification of the Spline functions does not relocate too significantly the end points of the arteries, it may happen that the extremity of the branch located onto the bifurcation may be slightly shifted away from the other two arteries; in other words, at the bifurcation node, any of the three arteries might not connect any longer with the others, we deal with this issue by simply locating the new extremity coordinates and by shifting the whole set of curve coordinates back toward the center of the 3D crop. 

Besides the convolution with a spherical kernel, that is needed to set the artery's thickness, our method allows to set a target grey level amplitude. We simply multiply the binary envelope of the modeled vasculature by the desired target grey level. Moreover, for the thickened artery to have a realistic shape, it is important that the convolution kernel is not perfectly spherical. Indeed, if a sphere was being applied, the so-obtained modeled artery would take the form of a perfect tube, which hardly happens on real-life arteries. Hence, we apply a geometric distortion (elastic deformation) to the 3D kernel before applying the convolution. Furthermore, it is also possible to make the modeled artery inhomogeneous by applying various kernels (or more precisely various elastic deformations to the same kernel) along the vessel's centerline. 

\subsection{Modeling the surrounding brain matters}

Now that we have been through the details of the shape modeling, let us next present the second part of the synthetic model: the brain, composed of fluids and white/gray matters, all these, being affected by a reconstruction background noise.

\subsubsection{Collecting the target statistics }

As can be observed on Fig.~\ref{fig:VaMosProcess}, for the background generation (brain and noise), the cropped MRA-TOF patch first goes through a separation of the various brain components. 
In fact, based on the same rationale that was previously used (section~\ref{ssec:ModGeomArt}) to model the arteries, we can easily imagine to include the various brain matters within the model. Indeed, the cerebral arteries are surrounded by various matters, each one having a particular radio-opacity, \textsl{i.e.} a different gray level. The white/gray matter, when acquired through MRA appears with relatively high gray levels, the Cerebro-Spinal Fluids (CSF), the ventricle or the Corpus Callosum are commonly displayed with much lower luminances. 
However, unlike our previous study \citep{ICPR2022}, the areas of darker matters of the brain are no more randomly determined; we believe that those local low contrast shapes may be of paramount importance while modeling the 3D crops of our arteries and bifurcation, and hence we intend to include a faithful representation of  the fluids areas within the synthetic model. 
The separation between the darker and brighter matters (vasculature excluded) can either be performed via a simple multi-threshold segmentation \citep{Liao2001} or by using Gaussian Mixture Models (actually, our synthetic model allows both approaches).
Once located, each matter can then be geometrically distorted, before replicating it's overlaying noise.

\subsubsection{Noise generation}

When going through Gaussian blur, the input image $I(x,y)$ is filtered as shown below in eq.~\ref{eq:GBlur}.

\begin{equation}
O(x,y)=\sum_{i=-\infty}^{\infty}\sum_{j=-\infty}^{\infty}\frac{1}{2\pi\sigma_{G}^{2}}e^{-\frac{i^{2}+j^{2}}{2\sigma_{G}^{2}}}I(x+i,y+j)\label{eq:GBlur}
\end{equation}

The Bienaym\'e's identity states that 

\begin{equation}
\begin{array}{c}
Var\left(\sum_{i=1}^{n}X_{i}\right)=\\
\sum_{i=1}^{n}Var(X_{i})+\sum_{i,j=1,i\neq j}^{n}Cov(X_{i},X_{j})
\end{array}\label{eq:Bienayme}
\end{equation}

Thus, the variance of a linear combination is:

\begin{equation}
\begin{array}{c}
Var\left(\sum_{i=1}^{n}c_{i}X_{i}\right)=\\
\sum_{i=1}^{n}c_{i}^{2}Var(X_{i})+2\times\sum_{i,j=1,i\neq j}^{n}c_{i}c_{j}Cov(X_{i},X_{j})
\end{array}\label{eq:VarLinearCombination}
\end{equation}

However, if $X_{i},...,X_{n}$ are pairwise independent integrable random variables $\left(Cov(X_{i},X_{j})=0,\,\forall(i\ne j)\right)$, which we assume in the following, then:

\begin{equation}
Var\left(\sum_{i}c_{i}X_{i}\right)=\sum_{i}c_{i}^{2}Var(X_{i})\label{eq:VarIdentity2}
\end{equation}

where $c_{i}$ are constants. 

We consider that the variance of the input image is $Var\left[I(x+i,y+i)\right]=\sigma_{0}^{2}$, our goal here is to estimate the variance of the output (filtered) image $Var\left[O(x,y)\right]=\sigma_{f}^{2}$. Thus,

\begin{equation}
\sigma_{f}^{2}=\sigma_{0}^{2}\sum_{j=-\infty}^{\infty}\sum_{i=-\infty}^{\infty}\left(\frac{1}{2\pi\sigma_{G}^{2}}e^{-\frac{i^{2}+j^{2}}{2\sigma_{G}^{2}}}\right)^{2}\label{eq:sigma}
\end{equation}

For large $\sigma_{G}$, the squared Gaussian is smooth and the sum can be approximated as:

\begin{equation}
\begin{array}{c}
\sigma_{f}^{2}\approx\sigma_{0}^{2}\int_{-\infty}^{\infty}\int_{-\infty}^{\infty}\left(\frac{1}{2\pi\sigma_{G}^{2}}e^{-\frac{i^{2}+j^{2}}{2\sigma_{G}^{2}}}\right)^{2}di.dj\\
=\frac{\sigma_{0}^{2}}{4\pi\sigma_{G}^{2}}
\end{array}\label{eq:sigmaf_sigma0}
\end{equation}

and thus, 

\begin{equation}
\sigma_{f}\approx\frac{\sigma_{0}}{2\sigma_{G}\sqrt{\pi}}\label{eq:sigmaf_sigma0-2}
\end{equation}

In summary, when an image composed of Gaussian noise of standard deviation $\sigma_{0}$ is being filtered by a Gaussian filter of standard deviation $\sigma_{G}$, the so-obtained filtered image has a standard deviation of $\sigma_{f}$ according to the eq.~\ref{eq:sigmaf_sigma0-2}.

However, for our particular purpose, we intend to determine which Gaussian filter (of standard deviation $\sigma_{G}$) shall be used on the input image so as to obtain a filtered image with a given target statistics ($\sigma_{f}$), and hence $\sigma_{G}\approx\sigma_{0}/(2\sigma_{f}\sqrt{\pi})$.

The process starts thus with the generation of a high frequency Gaussian Noise of average set to our target 3D crop. This noise will then be smoothed out using a Gaussian filter of standard deviation $\sigma_{G}$. The resulting image (of standard deviation $\sigma_{f}$) will thus present strong similarities with the target portion of the MRA-TOF being modeled. 

The performances evaluation could be conducted in several ways. The generated noise could be considered as a texture, and the Haralick features \citep{Haralick1979} might be exploited to measure the correspondence between the modeled patch and its corresponding ground truth. Similarly, concerning the noise's spatial frequencies, the Variance of Laplacian, or the Tenengrad coefficient could help assessing a blur factor \citep{Ali2018}. 
Interested reader may refer to~\citep{ICPR2022} for details. 
In this study, we do not make any attempt to evaluate the modeled texture, we expect that an accurate modeling will inevitably lead to high aneurysm detection performances. Hence, the model accuracy will inevitably be strongly linked to the CNNs accuracy (in terms of both ICA segmentation and detection).

\subsection{Modeling and adjoining the aneurysm}

Once the bifurcation has properly been modeled, a synthetic aneurysm can finally be incorporated. 
A simple 3D sphere is first created and then distorted using elastic deformations. Further, the ICA center is aligned onto the bisector between the two daughter arteries. The distance from the aneurysm center to the bifurcation node has been computed as shown in eq.~\ref{eq:Distance_ICA}.
\begin{equation}
\mathcal{D} = r \times \gamma +\sqrt{\left(\frac{R}{tan(\Theta/2)}\right)^2+R^2}
\label{eq:Distance_ICA}
\end{equation}
where $r$ is the aneurysm radius, $R$ is the average radius of the branches forming the bifurcation, and $\Theta$ stands for the angle formed by the two daughter arteries. Thanks to a growth parameter $(\gamma)$, we can automatically adjust the shift from the aneurysm center and the vessel wall where the daughter arteries split. Hence, we can model various states of growth for a given aneurysm.

\subsection{Model examples}

\begin{figure}[!ht]
\begin{centering}
\includegraphics[width=0.63\columnwidth]{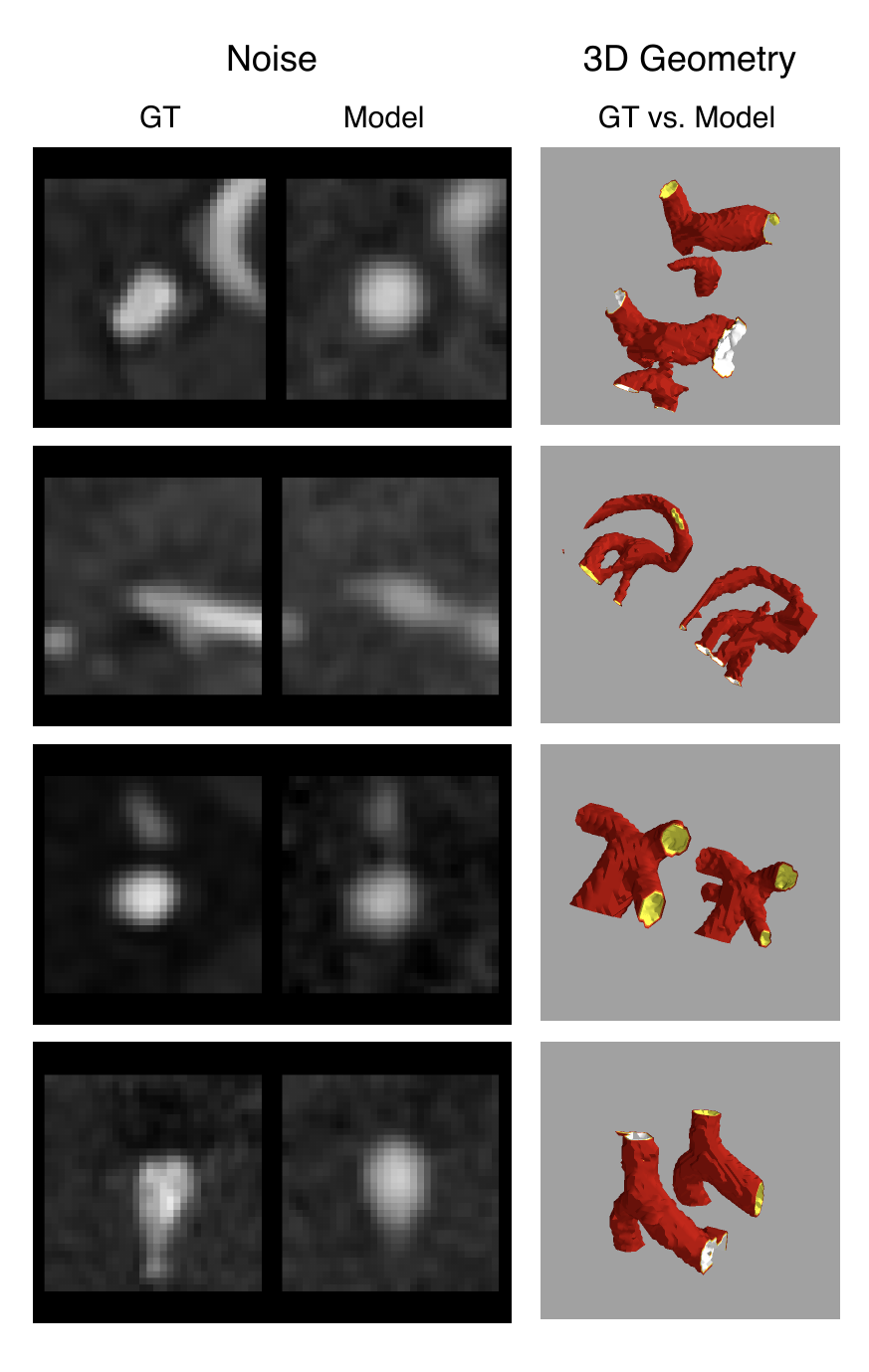} 
\par\end{centering}
\caption{Comparison between the modeled bifurcations and the Ground Truth crop from a MRA-TOF. 
We show the comparison on terms of both gray level voxels patches (leftmost panels) and 3D bifurcation layout (rightmost panels). \label{Comp_VaMos_GT_Bif}}
\end{figure}

Concerning the bifurcation model, Fig.~\ref{Comp_VaMos_GT_Bif} shows, for four different extracted bifurcation patches, a comparison between the Ground Truth (GT) crop and the modeled patch. Images on the left panel shows a 2D slice of the gray level voxels for both the Ground Truth and the Model, whereas the rightmost images represent the 3D layouts of the arteries.
We can notice that both the geometrical configuration of the bifurcations and the gray level distribution seem to be rather nicely modeled and mimic very accurately the TOF patch. Indeed, a very wide variety of bifurcations (no matter how complex the shapes are) can easily be modeled. We can notice that some subtle diameter or tortuosity modifications are faithfully brought on the modeled bifurcations.

\begin{figure}[!ht]
\begin{centering}
\includegraphics[width=0.63\columnwidth]{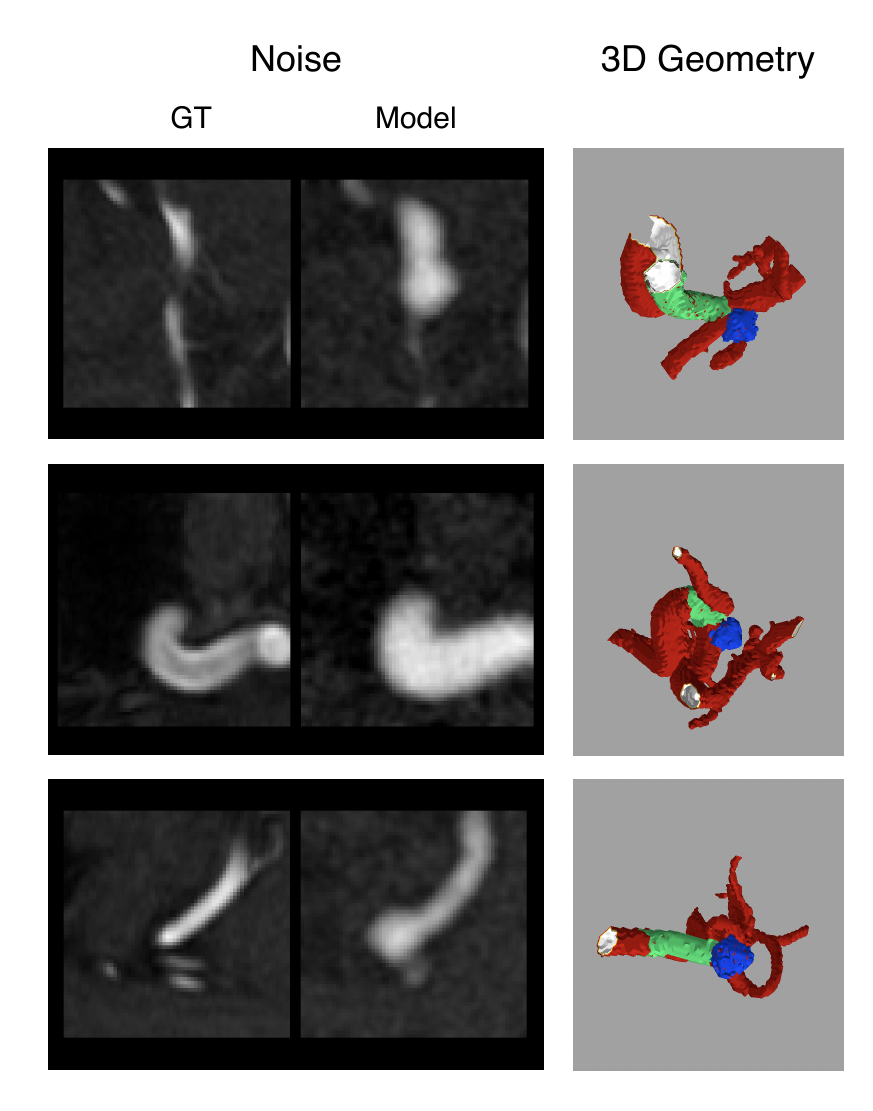}
\par\end{centering}
\caption{Comparison between the modeled bifurcations (bearing an aneurysm). In the upper sub-figures (3D representations), the aneurysm is represented in blue, the mother artery in green. \label{Comp_VaMos_GT_ICA}}
\end{figure}

Besides the bifurcations themselves, it is crucial for the aneurysm to be accurately modeled and most importantly well positioned onto the bifurcation artery wall. We present three different examples on Fig.~\ref{Comp_VaMos_GT_ICA}. Again we can observe how the slice gray levels are faithfully represented in the model. On the rightmost images, the aneurysms are represented in blue, whereas the mother artery is depicted in green. 

So far, the synthetic aneurysm model can only reproduce the unruptured and untreated aneurysm. Indeed, ruptured or treated aneurysm presents significant differences, treated aneurysms are radiolucent (darker voxels), and can easily be mistaken by surrounding fluids,  whereas ruptured aneurysms present very different shapes (higher order spherical aberrations, larger elongation, etc.) within the MRA-TOF images. 
Of course, we could imagine modulating the ICA gray level amplitude and shape so as to model ruptured/treated aneurysms. 
Such a study might be the topic of a future research project.

\section{Experimental Results}
\label{sec:ExpRes}

The sole purpose of this synthetic model is actually to be able to effortlessly build up significant images datasets in order to efficiently train Convolutional Neural Networks for pattern recognition tasks. 

This section is devoted to the experimental results related to the effectiveness of our synthetic model for the development of a fully automated Deep Learning (DL) network aiming to detect intracranial aneurysms (ICAs). Specifically, synthetic aneurysms are incorporated into Magnetic Resonance Angiography (MRA) Time-Of-Flight (TOF) scans that are originally aneurysms-free. Embedding such synthetic aneurysms, along with their corresponding labels, enables a much faster generation of training data compared to the time-consuming process of annotating real aneurysms.

\begin{figure*}[!ht]
\begin{centering}
\includegraphics[width=\textwidth]{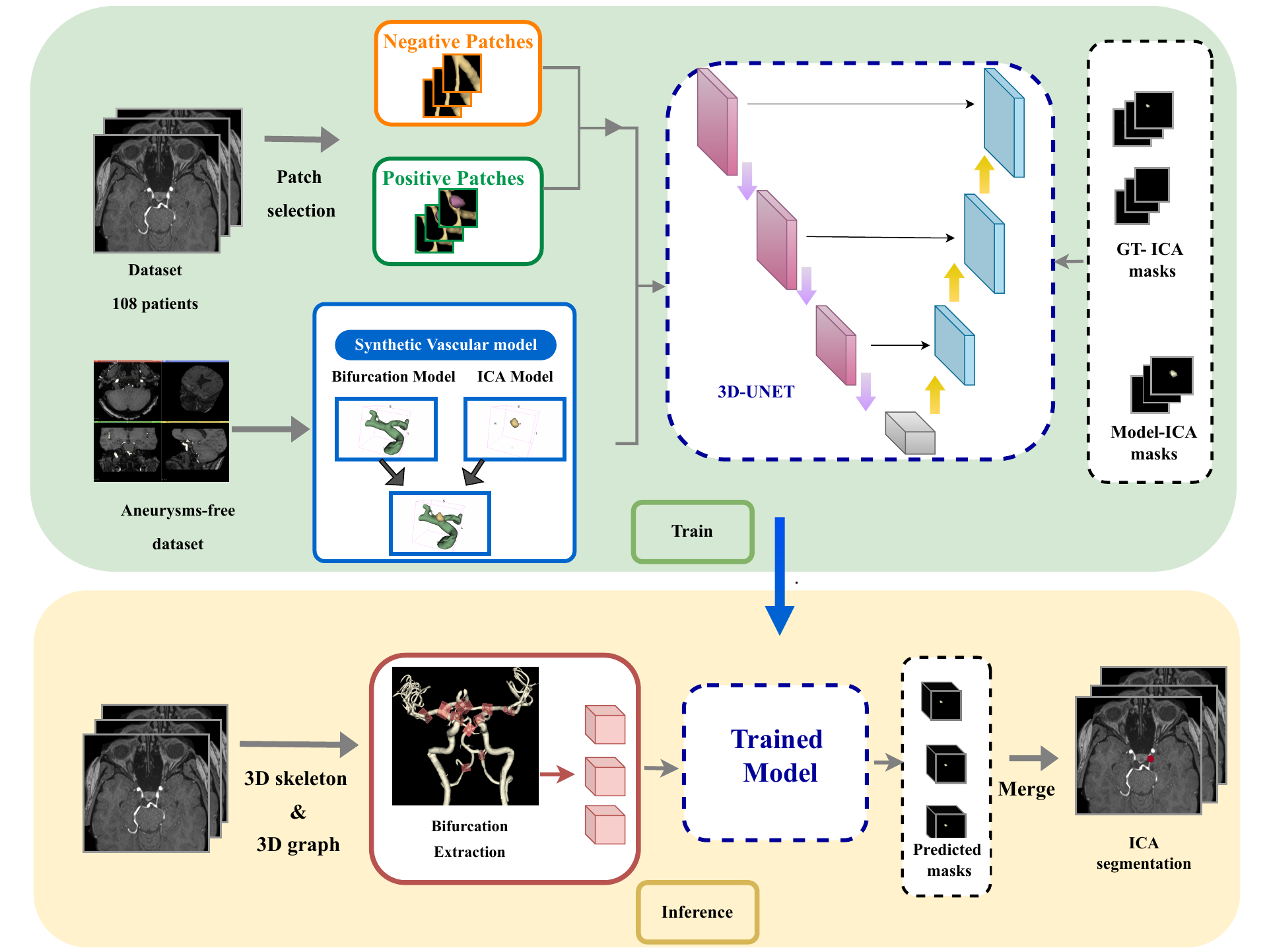}
\par\end{centering}
\caption{Overview of the global procedure, encompassing the training step using the synthetic images and the inference step. \label{fig:GlobalProcess}}
\end{figure*}

Fig.~\ref{fig:GlobalProcess} shows an overview of the entire aneurysm detection process. The U-Net is trained on both the synthetically generated images and some actual MRA-TOF acquired from patients with aneurysm. The trained U-Net model is then applied onto some extracted bifurcations during the inference phase. 

We will first introduce the Deep Learning based  method being used, along with the corresponding dataset, the training strategy and the evaluation approach. Finally, we will evaluate the efficiency of our approach, and estimate the benefit of adjoining synthetic images for the CNN training step.

\subsection {DL based detection of intracranial aneurysms}

In this work, we use a 3D Convolutional Neural Network (CNN) segmentation model as the deep-learning algorithm for the automated detection of intracranial aneurysms on MRA-TOFs.

\subsubsection{Dataset}
For this study, a total of 190 MRA-TOFs scans of unruptured intracranial aneurysms were collected from more than thirty different French institutions (some DICOM headers did not contain the institution name, marked as unknown). These images were randomly divided into two datasets: a training set comprising 108 images, used for training and validation, and a separate test set containing 82 images.
Furthermore, 14 additional TOF images not containing any aneurysm were included in the test set to evaluate the performances of the developed model. These aneurysm-free images were selected to assess the model's ability to accurately distinguish between aneurysm and no-aneurysm cases. Out of these 190 subjects, 58 had more than one aneurysm. 
Therefore, overall 254 aneurysms are included in this study. Tables \ref{tab2} and \ref{tab3} show their respective sizes and locations.

From now on, when evaluating the detection performances per bifurcation labels, we will group various labels altogether. Indeed, we have grouped the three bifurcations along the basilar artery together (M, N and O), as overall, few aneurysms pop out along the anterior portion of the Circle of Willis. Moreover, due to the anatomical configuration of the CoW on humans, both the PCoA and MCA arteries commonly reach the Internal Carotid Artery in the same vicinity, and hence, the bifurcations E (\textit{resp.} F) are very close to the bifurcations I (\textit{resp.} J), and thus, the aneurysms emerge in a very close neighborhood (possibly within the same cropped area encompassing two bifurcations of interest).

\begin{table}[!ht]
\caption{Aneurysms radii in the training and test data sets.\label{tab2}}
\begin{center} 
\begin{tabular}{|c|c|c|}
\hline
\textbf{Aneurysm radius }&\textbf{Train}&\textbf{Test}\\
\hline
$\leq 2$ mm& 19 & 47\\
\hline
2-3 mm &64&69\\
\hline
$>3$ mm &44&11\\
\hline
\end{tabular}
\end{center}
\end{table}

\begin{table}[!ht]
\caption{Number of aneurysms based on their location within the CoW. \label{tab3}}
\begin{center} 
\begin{tabular}{|c|c|c|}
\hline
\textbf{Bifurcation label}&\textbf{Training set}&\textbf{Testing set}\\
\hline
A-B &11&17\\
\hline
C-D &1&6\\
\hline
E-F-I-J &12&18\\
\hline
G-H &54&44\\
\hline
K-L &48&37\\
\hline
M-N-O &1&5\\
\hline
Total & 127 & 127 \\
\hline
\end{tabular}
\end{center}
\end{table}

A trained operator performed the annotations to build up the dataset. Subsequently, a neuroradiologist with 10 years of experience,  carefully reviewed the cases to ensure the exclusion of any potential false positives or false negatives that might have been initially reported in the original annotation.

\subsubsection{Real patches selection and Neural network}

In our study, we have used a patch-based approach for the aneurysm detection process. Instead of using entire volumes, we feed the neural network with 3D patches of size $64 \times 64 \times 64$. 

To select the 3D training patches, we employed a random extraction strategy. Specifically, for each aneurysm, we randomly extract 10 patches centered around the vicinity of the aneurysm (random shifts along the $x$, $y$ and $z$ directions). However, for each extracted patch, we ensure that the entirety of the ICA is included within the cropped area. 
This approach ensures that the training dataset contains diverse samples representing different locations of the aneurysm within the extracted patches. 
For negative samples, we simply extract some 3D patches encompassing some cerebral arteries, but without any aneurysm. For each  MRA-TOF volume, we have extracted 20 such patches. 
This selection strategy allows for a comprehensive representation of both positive (presence of an aneurysm) and negative (aneurysm-free artery) samples, and hence, ensures the generation of a training dataset capturing the necessary characteristics for an effective model learning.

We have opted for a segmentation network for the dual tasks of \textit{i)} ICA mask segmentation and \textit{ii)} the subsequent ICA detection. 
The segmentation process uses a 3D U-Net architecture \citep{cicek20163d}. The U-Net architecture follows an encoder/decoder structure. Each level of the encoder consists of convolution layers, and max pooling layers. We used upsampling layers in the decoding branch instead of transpose convolutions. The ReLU activation function was applied to all layers, except for the final layer, which was followed by a sigmoid function.
To train the model, we applied the Adam optimization algorithm with a learning rate of 0.0001 and a batch size of 16. The Combo loss function \citep{Taghanaki2019} which combines both the Dice coefficient and the binary Cross-entropy loss was used in conjunction with this optimization algorithm.

\subsubsection{Data Augmentation}
Overall, 134 MRA-TOFs free of any aneurysm were collected from the previous work described in \citep{NADER2023102919}. 
These scans were utilized for constructing 3D synthetic cropped portions mimicking the characteristics of an original MRA-TOF, as explained in the previous section. For this purpose, a total of 998 synthetic patches were modeled. Each patch was specifically centered on the bifurcations of interest within the CoW. 
Once the bifurcation has been accurately modeled, a synthetic aneurysm is incorporated thanks to the 3D model. This is achieved by varying the radius parameter and by applying degrees of elastic deformations to simulate the diverse characteristics observed in actual aneurysms. 
Tables \ref{tab4} and \ref{tab5} show the distribution of the modeled aneurysms with respect to their locations and sizes.

\begin{table}[!ht]
\caption{Number of modeled patches for each bifurcation label. \label{tab4}}
\begin{center} 
\begin{tabular}{|c|c|}
\hline
\textbf{Bifurcation label}&\textbf{\# of ICAs}\\
\hline
A-B &165\\
\hline
C-D &158\\
\hline
E-F &156\\
\hline
G-H &175\\
\hline
I-J &102\\
\hline
K-L &111\\
\hline
M-N-O &131\\
\hline
\end{tabular}
\end{center}
\end{table}

\begin{table}[!ht]
\caption{Number of modeled aneurysms per radius range.\label{tab5} }
\begin{center} 
\begin{tabular}{|c|c|}
\hline
\textbf{Radius}&\textbf{Count}\\
\hline
$\leq 2$ mm& 292\\
\hline
2-3 mm &596\\
\hline
$>3 $mm &110\\
\hline
\end{tabular}
\end{center}
\end{table}

\subsubsection{Evaluation approach}
\label{ssec:Eval}

To validate the possible improvements brought by the use of the synthetic model, we have conducted two separate experiments.
In the first experiment (\textit{Exp.\#1}), we have trained a baseline model using actual MRA-TOFs patches. To assess the performance of our model, we employed a four fold cross-validation approach. The dataset was split into four folds with each fold containing 27 samples. 
During each cross-validation split, three folds (81 samples) were used to train the model, while the remaining fold (27 samples) was reserved for validation purposes and for hyperparameter optimization. This process was repeated four times, to ensure that all 108 TOFs were ultimately used for evaluation. In the second experiment (\textit{Exp.\#2}), we trained another segmentation network but augmenting the training dataset with 998 synthetic patches. Similarly, we evaluated the performance of the model using the same cross-validation split. Following the training phase, the model evaluation employs the holdout test set using the four-fold models. Subsequently, the resulting predictions from these four models were averaged to derive the final predictions.

In the inference phase, for both experiments, we adopt a prior anatomical selection of patches. We only retain the patches being centered onto some bifurcations of the cerebral vasculature.
By focusing on patches centered around the cerebral artery bifurcations, the inference process aims to target the regions being most susceptible to witness an aneurysm development. This approach is based  on the anatomical knowledge of aneurysm occurrence, enhancing the accuracy of the results. Hence, as previously explained, to select the corresponding patches, an automated vessel segmentation step was performed using a pre-trained U-Net segmentation algorithm. 
The details of the specific pre-trained  network can be found in \citep{NADER2023102919}.  
Then, a 3D undirected graph \citep{Bifurc_patent_2018,Nouri2020} is generated from the extracted skeleton to extract the corresponding bifurcations.

For both experiments, we have analyzed the patient-level sensitivity, the lesion-level sensitivity and the false positive rate (number of false positive per TOF). The sensitivity quantifies the proportion of true aneurysms correctly identified by our method, thereby measuring its ability to capture actual aneurysms while the false positive count per TOF provides insights about  the method  possibly producing an excessive number of false identifications, which could lead to an unnecessary burden for the neuroradiologists. To compute the detection performance evaluation metrics, we consider each segmented connected component issued from the binary output of the U-Net as a potentially detected object. Each connected component whose center of mass falls no farther away from the maximum radius of a true aneurysm mask, is thus considered as a true positive detection. Otherwise, it is regarded as a false positive detection. 

In addition, we applied the evaluation protocol for assessing the segmentation performance of true (ICAs) as described in the ADAM challenge \citep{Timmins2021}. This protocol focuses on evaluating the segmentation metrics only for the true detected ICAs, excluding any false positives, to simulate how the tool could be practically used by neuroradiologists.

\subsection{Performance analysis}

Let us now examine the performances of our proposed method on the test set, in terms of global detection rate as well as on a per-bifurcation scenario. We will provide an evaluation of both the detection ability and the segmentation accuracy. 

\subsubsection{Overall detection  performance}
\begin{figure}[!ht]
\begin{centering}
\includegraphics[width=0.35\textwidth]{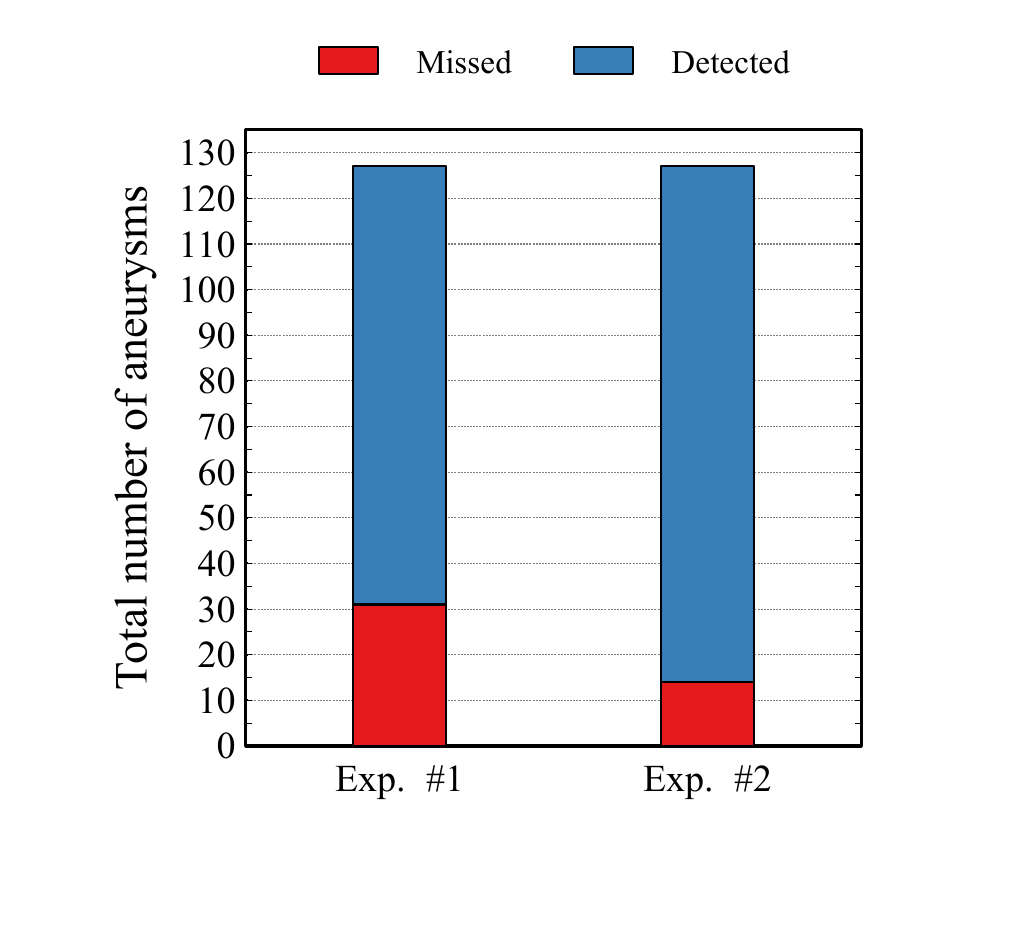}
\par\end{centering}
\caption{Detection performance of the CNN using real data vs real and synthetic data\label{detect_perf}}
\end{figure}

As previously mentioned, the study compares the performances of two CNN training approaches. In \textit{Exp.\#1}, the CNN is exclusively trained onto real data. In contrast, \textit{Exp.\#2} involves training the same CNN using a combination of real data patches and synthetic data patches. The overall detection performance is shown in Fig.~\ref{detect_perf}.

During \textit{Exp.\#1}, the CNN successfully identified a total of 96 aneurysms within a dataset containing 127 instances, which corresponds to a lesion-level sensitivity of $75.60\%$. Notably, this sensitivity further improved to $88.97\%$ for \textit{Exp.\#2} with 113 detected aneurysms. 
In \textit{Exp.\#1}, the patient-level sensitivity reached $79.65\%$, indicating the ability to correctly identify aneurysms at the patient level. Meanwhile, \textit{Exp.\#2} attained a  higher patient-level sensitivity of $91.36\%$.

\subsubsection{False detections}
These results show a high diagnosis performance, with lesion-wise sensitivity notably improving by incorporating synthetic patches, reaching $89\%$ on the  test set. The synthetic data proved useful as a complementary tool to reduce the missed aneurysms rate. However, obtaining a high sensitivity may unfortunately lead to a slightly higher false-positive detection rate.
Overall, the network exhibits an average false positive rate of $0.22$ in \textit{Exp.\#1}, whereas when incorporating synthetic patches, a marginal increase in this false positive rate is observed (reaching $0.40$). 
The slight increase in the false positive rate upon integrating synthetic patches, along with the significant  sensitivity gain, emphasizes the importance of adopting \textit{Exp.\#2}. This illustrates the trade-off between sensitivity and false positive rate and highlights the added value of the data augmentation.

Based on a thorough visual inspection of the false-positive detections, we have identified various reasons for these FP, including \textit{1)} The complex anatomy of the internal carotid artery, with sometimes rather strong variations in vessel diameter, a high tortuosity, and a very significant bending, right below the ophthalmic artery that can be confused with a large aneurysm ;
\textit{2)} Brain arteries are susceptible to flow-related changes due to factors such as atherosclerosis (calcified plaque) or stenosis. These conditions can alter blood flow patterns and vessel appearance, potentially leading aneurysm-like vessels shapes, and hence, to false-positives.
\textit{3)} At the emergence of a daughter artery on a bifurcation, at the very basis of the artery, an outpouching can be formed. In other words, the artery stars with a conic shape, exhibiting a broad base located at the bifurcation. This is clinically referred to as an \textit{infundibulum}. Sometimes, such uncommon shapes can be mistakenly detected as being an aneurysm.
\textit{4)} Our vascular model has been designed in such a way to model a thrombosis effect within the aneurysms. Indeed, we have noticed that for the larger circular aneurysms, a thrombosis often appears nearby the ICA center; the blood flow may circulate along the aneurysm walls, forming some sort of vortex, and hence, leading to a slower blood displacement toward the center (inducing a more radio-opaque area). Unfortunately, such a phenomenon can also occur along the Internal Carotid Artery, thus leading to false positive detections.

\subsubsection{Impact of aneurysms size and locations}
The performance variability when considering aneurysm size, is depicted in Table \ref{tab11}. \textit{Exp.\#1} yielded a detection rate of $0.5106$ for very small aneurysms (less than 2 mm), while \textit{Exp.\#2} achieved an improved rate of $0.7659$. 
The CNN was able to detect a greater proportion of aneurysms falling  within the size range of 2 mm to 3 mm. This trend was observed in both experiments, with a notable increase observed for \textit{Exp.\#2} ($0.8840$ versus $0.9565$). Both methodologies, however, demonstrated a  detection rate of 1 for aneurysms with a radius exceeding 3 mm.

\begin{table}[!ht]
\caption{Lesion-level sensitivity according to the aneurysm size for the test set \label{tab11}}
\begin{center} 
\begin{tabular}{|c|c|c|}
\hline
\textbf{Radius}&\textbf{\textit{Exp.\#1}}&\textbf{\textit{Exp.\#2}}\\
\hline
$\leq 2$ mm& 51.06 $\%$ & 76.59 $\%$\\
\hline
2-3 mm &88.40 $\%$&95.65 $\%$\\
\hline
$>3$ mm &100 $\%$&100 $\%$\\
\hline
\end{tabular}
\end{center}
\end{table}

\begin{figure}[!ht]
\begin{centering}
\includegraphics[width=0.85\columnwidth]{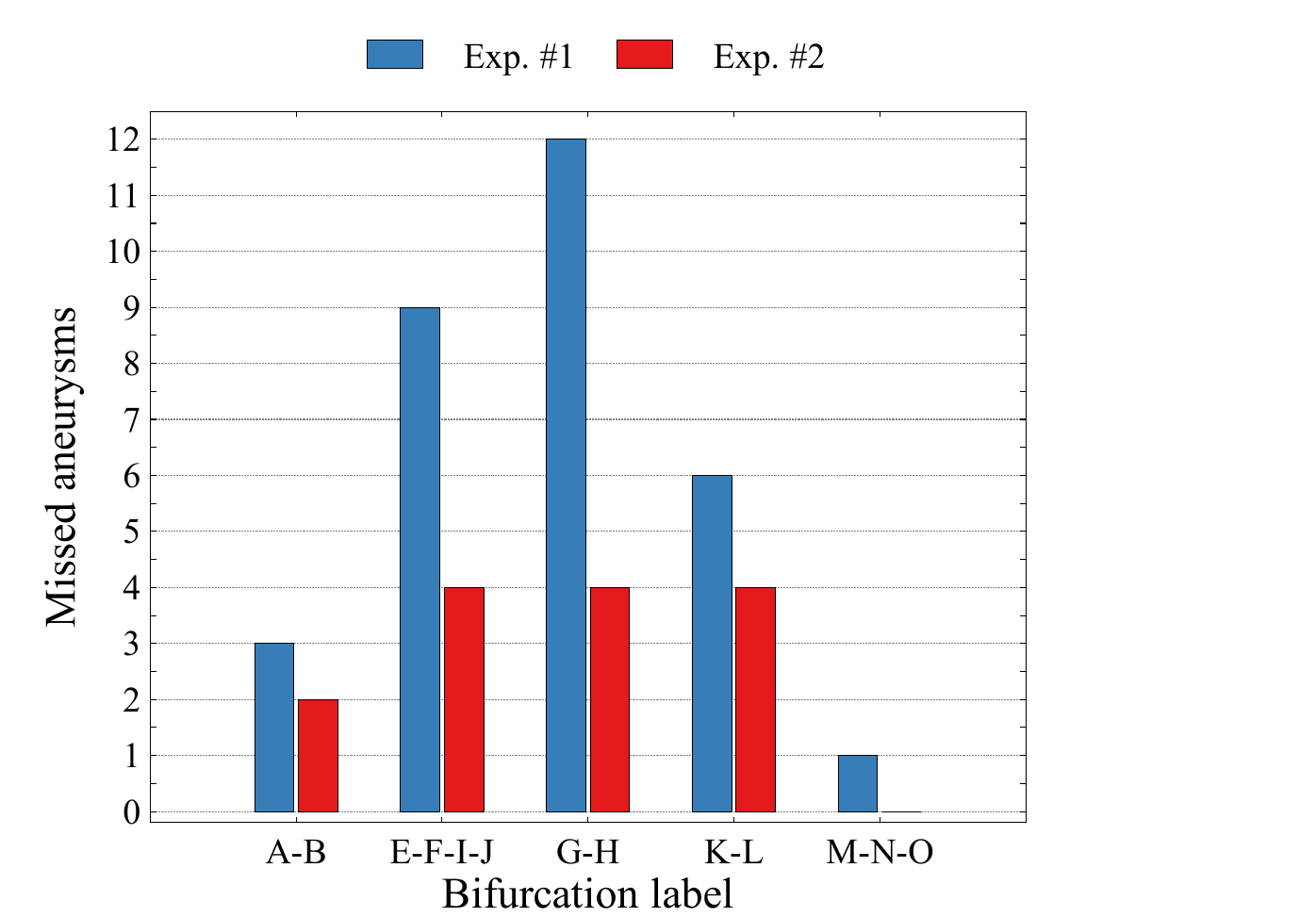}
\par\end{centering}
\caption{Missed detections with respect to the aneurysms positions in the test dataset\label{fig10}}
\end{figure}

Regarding the impact of aneurysm location, Fig.~\ref{fig10} depicts the number of missed aneurysms per bifurcation of interest across the test dataset. Specifically, in \textit{Exp.\#1}, the CNN missed a higher number of aneurysms located along the Middle Cerebral Artery (MCA) (G-H bifurcations), as well as along the Internal Carotid Artery, with a significant concentration in the bifurcation segment separating this artery into the smaller Ophthalmic Artery (OA) (labels K \& L). Additionally, missed aneurysms are observed within the branches of the Anterior Cerebral Artery (ACA) and the Posterior Communicating Artery (PCOM), which corresponds to bifurcations (E, F, I \& J).

In contrast, for \textit{Exp.\#2}, there is a substantial decrease in the number of missed aneurysms within the three locations cited  below.
The impact of the aneurysm location on the detection rates for both experiments is presented in Table \ref{tab12}.
One can notice on this Table that for each bifurcation label, the amount of collected aneurysms matches remarkably well the percentages of occurrences as previously described on Fig.~\ref{fig:SchematicWillis}. 
Hence anatomically, we are able to collect fewer aneurysms onto certain bifurcations. 
Moreover, it is important to highlight that fewer synthetic aneurysms can be modeled onto the bifurcations labeled C, D, N and O (\textit{c.f.}, Table~\ref{tab4}) as, quite often, during the angiography exam, these bifurcations (at the farther ends of the MRA 3D stack) may be cropped out of the acquisition area.

\begin{table}[!ht]
\small
\caption{The detection rates with respect to aneurysms location \label{tab12}}
\begin{center} 
\begin{tabular}{|c|c|c|c|}
\hline
\textbf{Label}&\textbf{Count}&\multicolumn{2}{c| }{\textbf{Detection Rate} ($\%$)} \\
 & & \textbf{\textit{Exp.\#1}} & \textbf{\textit{Exp.\#2}} \\ 
 \hline
\textbf{A-B}&17&82.35&88.23\\ 
\hline
\textbf{C-D}&6&100.00&100.00\\
\hline
\textbf{E-F-I-J}&18&50.00&77.77\\
\hline
\textbf{G-H}&44&76.19&90.90\\
\hline
\textbf{K-L}&37&83.78&89.18\\
\hline
\textbf{M-N-O}&5&80.00&100.00\\
\hline
\end{tabular}
\end{center}
\end{table}

\subsubsection{Segmentation performance of true ICA}

\begin{figure}[!ht]
    \centering
    \includegraphics[width=0.6\columnwidth]{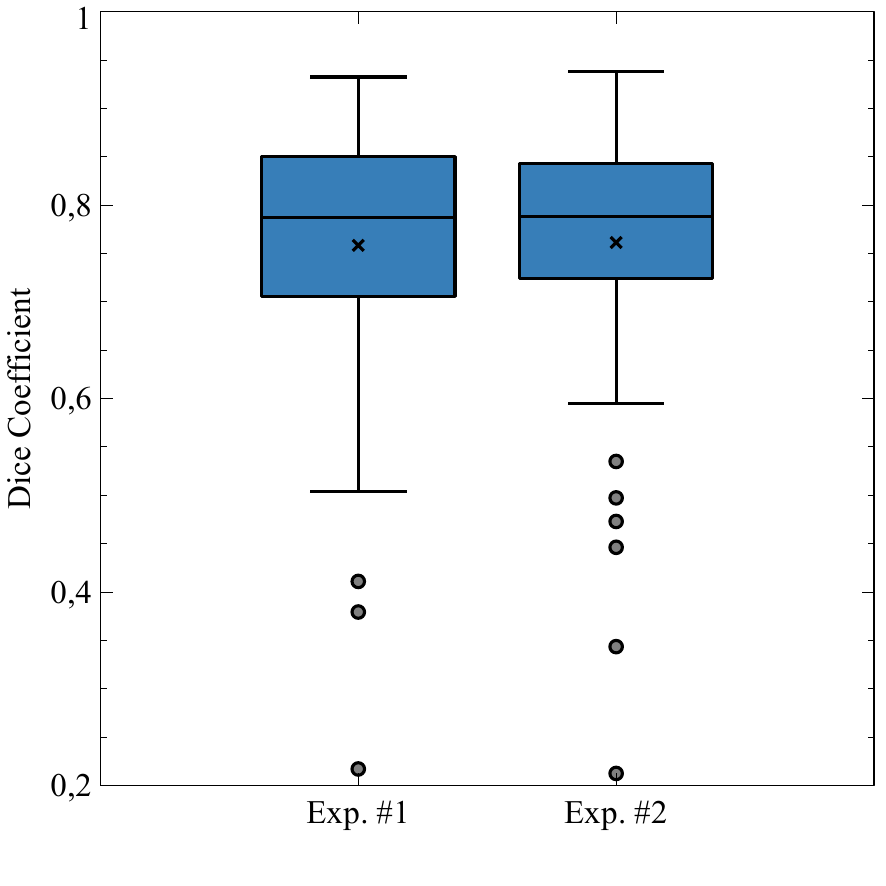}
\caption{\label{fig14} Dice similarity coefficient  of true ICAs }
\end{figure}

Fig.~\ref{fig14} displays the distribution of Dice score for the detected aneurysms within both experiments \#1 and \#2.  The average Dice score for \textit{Exp.\#1}  is 0.7585 $(\pm  0.13$) which indicates a fairly good similarity with the actual aneurysms masks. \textit{Exp.\#2}, achieves a  comparable Dice score of 0.7613 ($\pm 0.12$). It is important to note that a direct comparison between the two Dice coefficients is to be considered carefully, as the number of detected aneurysms differs between the two experiments. 

However, the visual analysis of the bar plots representing the segmentation performance for \textit{Exp.\#1} and \textit{Exp.\#2} reveals similarities in their overall appearance. A closer examination identifies specific outliers in \textit{Exp.\#2}, characterized by notably lower Dice scores. These outliers can be attributed to the detection of small aneurysms, a distinction not present in \textit{Exp.\#1}.

\section{Discussion and Conclusion}
\label{sec:Discussion}

In this section, we analyze the contribution brought by the synthetic vasculature model, being able to faithfully replicate (while cleverly altering) portions of MRA-TOF images.
Indeed, the synthetic model is composed of various processes, including a meticulous modeling of the cerebral arteries and bifurcations geometry, alongside the introduction of surrounding noise and finally, embedding aneurysms of various sizes and shapes. 
Our goal is to provide a substantial dataset that may improve the performances of several deep learning tasks including the segmentation or detection of the cerebral aneurysms. 
A salient highlight of our work is the successful generation of synthetic aneurysms with varying sizes, shapes, and locations. These artificial aneurysmal sacs have been integrated into modeled MRA scans originally lacking any aneurysm, thus resulting in an augmented dataset that aligns more closely with real-world scenarios. The important focal point of this approach is the strategic positioning of cerebral aneurysms within the Circle of Willis, enhancing the fidelity of the simulated model.

\begin{table*}[!ht]
\small
\caption{ Studies applying deep learning models on MRA for the automatic detection/segmentation of cerebral aneurysms\label{tabsum}}
\begin{center} 
\begin{tabular}{c|c|c|c}
\multirow{2}{*}{\textbf{Study}} & \multirow{2}{*}{\textbf{\#Subjects / \#aneurysms}} & \textbf{Lesion-wise} & \multirow{2}{*}{\textbf{FP/case}}\tabularnewline
 &  & \textbf{sensitivity (\%)} & \tabularnewline
\hline
\hline
\citep{nakao2018}&100 subj. / 104 ICAs (2D MIP patches)  &94.2&2.9\\
\hline
\multirow{2}{*}{\citep{Ueda2019}} & 521 subj. / 649 ICAs (Internal test set)& 91& $---$ \\
 &67 subj. / 80 ICAs ( External test set) &93& 5 \\           
\hline
\citep{stember2019}&86 subj. (2D MIP patches) &98.8& $---$ \\
\hline
\citep{Sichtermann2019}&85 subj. / 115 ICAs &90&6.1 \\
\hline
\multirow{2}{*}{\citep{joo2020}}&170 subj. / 147 ICAs  (Internal test set)&87.1& 0.005\\
&106 subj. / 63 ICAs (External test set)  &85.7&0.004\\
\hline
\citep{chen2020}&35 subj. / 35 ICAs &82.9& 0.86 \\
\hline
\citep{Timmins2021}&141 subj. &67& 0.13  \\
\hline
\citep{di2023}&284 subj. / 198 ICAs &83&0.80 \\
\hline
\citep{ham2023}&15 subj. & 88.2 & 0.305 \\
\hline
OURS- \textit{Exp.\#1}&96 subj. / 127 ICAs&75.6 &0.22\\
\hline
OURS- \textit{Exp.\#2}&96 subj. / 127 ICAs&89 &0.40\\
\hline
OURS- \textit{Exp.\#3}&96 subj. / 127 ICAs &82 &2.3\\
\end{tabular}
\end{center}
\end{table*}

As the main finding of this study, the CNN trained using a combination of both genuine and synthetic patches led to a significantly improved sensitivity in detecting intracranial aneurysms compared to a CNN trained solely on MRA-TOF data. While the latter missed $24.4\%$ of the lesions on the test data, including the synthetic patches during the training step significantly improved the ICA detection performances of the CNN. 
Indeed, the CNN missed a smaller portion of aneurysms with only $11\%$ on the test set. Only 14 aneurysms were missed in \textit{Exp.\#2}: eleven were tiny aneurysms, two exhibited uncommon shapes (high order spherical aberrations), and finally, one was presenting a shape strongly similar to an infundibulum (no clearly delineated aneurysm neck).

This research has successfully achieved a high level of detection sensitivity, showcasing its potential as supplementary tool for neuroradiologists to address the issue of overlooked aneurysms. 
Nevertheless, it is important to note that the integration of synthetic patches, while effective in boosting sensitivity, can potentially contribute to a slightly increased false-positive rate. 
This issue might arise if the synthetic patches actually introduce irregular aneurysms shapes that do not very faithfully reflect the genuine aneurysm sacs as acquired on MRA-TOFs. However, the training methodology, employed alongside with a judicious selection of the synthetic model parameters has yielded an improved  detection performance while maintaining a low rate of false-positive detections. Specifically, the false-positive detection rate stood at 0.40 within \textit{Exp.\#2}, in contrast to 0.22 in \textit{Exp.\#1}.
The ability to maintain a minimal count of false-positive detection rate can also be explained by 2 factors: 
To begin with, the use of a prior selection of patches (3D undirected graph generated from the skeleton, as explained in section~\ref{ssec:Eval}) during the inference phase by extracting patches around vascular bifurcations. Consequently, this approach minimizes the susceptibility of the algorithm to confuse non-vessel structures and reduces the likelihood of incorrect predictions.
Furthermore, the final prediction on the test set is derived by aggregating the probabilities obtained from 4 cross validation models. This strategy enhances the overall robustness of the model's predictions.
Moreover, the mean Dice score index of true ICAs is 0.76 which is relatively high. This suggests that this automatic segmentation method performs at a similar level to manual segmentation once the true ICA has been correctly identified. The automatic segmentation could save time and effort in the analysis of medical imaging data and potentially improve the efficiency in diagnosing and analyzing intracranial aneurysms.

Regarding the aneurysm size, our analysis demonstrated that there were no significant discrepancy in sensitivity for large aneurysms. Indeed, the sensitivity reached $100\%$ for aneurysms having a radius larger than 3 mm for both experiments. 
The CNN achieved an overall sensitivity of $51.06\%$ for detecting aneurysms being smaller than 2 mm in \textit{Exp.\#1}, which complies with common findings in state-of-the-art aneurysm detection studies. For such small aneurysms, the sensitivity may increase with the number of training cases involved. Notably, in \textit{Exp.\#2}, the sensitivity value increased to $76.59\% $ by training the model with small sizes synthetic aneurysms. Nevertheless, enhancing the diagnostic performance of the model in detecting small aneurysms may necessitate further exploration and training with a larger set of synthetic cases representing small aneurysms.

With regard to the sensitivity at different locations, the \textit{Exp.\#2} achieved better performances than \textit{Exp.\#1} for aneurysms situated onto the bifurcations G-H, E-F-I-J and K-L. It is important to note that this sensitivity increase appears to be more strongly linked to the size of aneurysms in these specific locations rather than solely on the location itself. In fact, the detection rates stratified according to aneurysm location and size were compared using  Fishers’ exact test. No significant difference was found between different locations (p-value=0.69).
However, the statistical analysis revealed a significant difference in aneurysm detection rates when stratified by different sizes (p-value=0.005). 

Indeed, it is important to interpret the sensitivity values for the categories C-D and M-N-O  with careful consideration. Particularly so, as the dataset comprises only few aneurysms in those locations compared to other locations. For future investigations, it could be valuable to explore the CNN efficiency with a larger population, \textit{i.e.}, incorporating more aneurysms located in regions such  M-N-O, and C-D. 

In addition to the retrospective study, it would be interesting to analyze the performances of a CNN being exclusively trained onto the synthetic patches. To this aim, a separate experiment was conducted. In this experiment (\textit{Exp.\#3}), the CNN was trained using only 998 patches generated from the synthetic model. The resulting outcome was then evaluated on the same test dataset. The findings demonstrated sensitivity results  higher than those from \textit{Exp.\#1}. Specifically, the sensitivity was recorded at $82\%$ for the test set (104 aneurysms detected). 
However, it is noteworthy that this increase in sensitivity was accompanied by a relatively elevated false positive rate, measuring 2.3 per case. Therefore, at the current stage, a deep learning segmentation network trained only onto the synthetic data cannot be clinically applied. Further improvements need to be brought to the vascular model to be exploitable as a standalone source of training images.

Table~\ref{tabsum} gives an idea of the overall performances of our method, as compared to various methods from the literature. 
Obviously, a direct comparison between the different methods is quite delicate, as the composition and size of the training/test image datasets can strongly differ. Moreover, a comparison should jointly consider the gaps in terms of sensitivity and false positive rate. The purpose of this Table is thus only to validate the performances of our proposed method, and demonstrate it can compete with various state-of-the-art approaches.

This study has been conducted on MRA-TOF acquisitions only, however, the model is, in its very nature, quite flexible and its adaptation to other modalities, such as CTA or DSA, should be relatively straightforward. The operating mode remains the same, only the background noise modeling might need some slight adjustments.

\section*{Source code availability}

The source code for the synthetic Vascular Models (VaMos) has been made available on a GitLab repository\footnote{\url{https://gitlab.univ-nantes.fr/autrusseau-f/vamos}}.

This source code has been initially developed with Python (version 3.9) on Mac OS X, and has been tested with Python versions 3.7 to 3.11 on both Mac OS and Linux (Mint and Ubuntu distributions).

\section*{acknowledgements}
This work was partially supported by the French ANR project “WECAN” (ANR-21-CE17-0006) and INSERM CoPoC \# MAT-PI-22155-A-01 (RVF23037NSA).

\bibliographystyle{model2-names.bst}\biboptions{authoryear}
\bibliography{references.bib}

\begin{thebibliography}{49}
\expandafter\ifx\csname natexlab\endcsname\relax\def\natexlab#1{#1}\fi
\providecommand{\url}[1]{\texttt{#1}}
\providecommand{\href}[2]{#2}
\providecommand{\path}[1]{#1}
\providecommand{\DOIprefix}{doi:}
\providecommand{\ArXivprefix}{arXiv:}
\providecommand{\URLprefix}{URL: }
\providecommand{\Pubmedprefix}{pmid:}
\providecommand{\doi}[1]{\href{http://dx.doi.org/#1}{\path{#1}}}
\providecommand{\Pubmed}[1]{\href{pmid:#1}{\path{#1}}}
\providecommand{\bibinfo}[2]{#2}
\ifx\xfnm\relax \def\xfnm[#1]{\unskip,\space#1}\fi
\bibitem[{Adams et~al.(1999)Adams, Laitt and Jackson}]{Adams99}
\bibinfo{author}{Adams, W.}, \bibinfo{author}{Laitt, R.},
  \bibinfo{author}{Jackson, A.}, \bibinfo{year}{1999}.
\newblock \bibinfo{title}{Time of flight 3d magnetic resonance angiography in
  the follow-up of coiled cerebral aneurysms}.
\newblock \bibinfo{journal}{Interv Neuroradiol.} \bibinfo{volume}{5},
  \bibinfo{pages}{127--37}.
\newblock \DOIprefix\doi{10.1177/159101999900500203}.
\bibitem[{Ali and Mahmood(2018)}]{Ali2018}
\bibinfo{author}{Ali, U.}, \bibinfo{author}{Mahmood, M.}, \bibinfo{year}{2018}.
\newblock \bibinfo{title}{Analysis of blur measure operators for single image
  blur segmentation}.
\newblock \bibinfo{journal}{Applied Sciences} \bibinfo{volume}{8},
  \bibinfo{pages}{807--839}.
\bibitem[{Arimura et~al.(2006)Arimura, Li, Korogi, Hirai, Katsuragawa,
  Yamashita, Tsuchiya and Doi}]{arimura2006computerized}
\bibinfo{author}{Arimura, H.}, \bibinfo{author}{Li, Q.},
  \bibinfo{author}{Korogi, Y.}, \bibinfo{author}{Hirai, T.},
  \bibinfo{author}{Katsuragawa, S.}, \bibinfo{author}{Yamashita, Y.},
  \bibinfo{author}{Tsuchiya, K.}, \bibinfo{author}{Doi, K.},
  \bibinfo{year}{2006}.
\newblock \bibinfo{title}{Computerized detection of intracranial aneurysms for
  three-dimensional mr angiography: Feature extraction of small protrusions
  based on a shape-based difference image technique}.
\newblock \bibinfo{journal}{Medical physics} \bibinfo{volume}{33},
  \bibinfo{pages}{394--401}.
\bibitem[{Autrusseau et~al.(2022)Autrusseau, Nader, Nouri, {L'Allinec} and
  Bourcier}]{ICPR2022}
\bibinfo{author}{Autrusseau, F.}, \bibinfo{author}{Nader, R.},
  \bibinfo{author}{Nouri, A.}, \bibinfo{author}{{L'Allinec}, V.},
  \bibinfo{author}{Bourcier, R.}, \bibinfo{year}{2022}.
\newblock \bibinfo{title}{Toward a 3d arterial tree bifurcation model for
  intra-cranial aneurysm detection and segmentation}, in:
  \bibinfo{booktitle}{IEEE International Conference on Pattern Recognition
  (ICPR)}, \bibinfo{address}{Montreal, QC, Canada}. pp.
  \bibinfo{pages}{4500--4506}.
\bibitem[{Baumgartner et~al.(2021)Baumgartner, J{\"a}ger, Isensee and
  Maier-Hein}]{baumgartner2021}
\bibinfo{author}{Baumgartner, M.}, \bibinfo{author}{J{\"a}ger, P.F.},
  \bibinfo{author}{Isensee, F.}, \bibinfo{author}{Maier-Hein, K.H.},
  \bibinfo{year}{2021}.
\newblock \bibinfo{title}{nndetection: a self-configuring method for medical
  object detection}, in: \bibinfo{booktitle}{Medical Image Computing and
  Computer Assisted Intervention--MICCAI 2021: 24th International Conference,
  Strasbourg, France, September 27--October 1, 2021, Proceedings, Part V 24},
  \bibinfo{organization}{Springer}. pp. \bibinfo{pages}{530--539}.
\bibitem[{Bogunovic et~al.(2013)Bogunovic, Pozo, Cardenes, Roman and
  Frangi}]{Bogunovic2013}
\bibinfo{author}{Bogunovic, H.}, \bibinfo{author}{Pozo, J.M.},
  \bibinfo{author}{Cardenes, R.}, \bibinfo{author}{Roman, L.S.},
  \bibinfo{author}{Frangi, A.F.}, \bibinfo{year}{2013}.
\newblock \bibinfo{title}{{Anatomical labeling of the circle of willis using
  maximum a posteriori probability estimation}}.
\newblock \bibinfo{journal}{IEEE Transactions on Medical Imaging}
  \bibinfo{volume}{32}, \bibinfo{pages}{1587--1599}.
\bibitem[{Bourcier et~al.(2017)Bourcier, Chatel, Bourcereau, Jouan, Marec,
  {Daumas-Duport}, {Sevin-Allouet}, Guillon, Roualdes, Riem, Isidor, Lebranchu,
  Connault, Tourneau, Gaignard, Loirand, Redon and Desal}]{Bourcier2017}
\bibinfo{author}{Bourcier, R.}, \bibinfo{author}{Chatel, S.},
  \bibinfo{author}{Bourcereau, E.}, \bibinfo{author}{Jouan, S.},
  \bibinfo{author}{Marec, H.}, \bibinfo{author}{{Daumas-Duport}, B.},
  \bibinfo{author}{{Sevin-Allouet}, M.}, \bibinfo{author}{Guillon, B.},
  \bibinfo{author}{Roualdes, V.}, \bibinfo{author}{Riem, T.},
  \bibinfo{author}{Isidor, B.}, \bibinfo{author}{Lebranchu, P.},
  \bibinfo{author}{Connault, J.}, \bibinfo{author}{Tourneau, T.},
  \bibinfo{author}{Gaignard, A.}, \bibinfo{author}{Loirand, G.},
  \bibinfo{author}{Redon, R.}, \bibinfo{author}{Desal, H.},
  \bibinfo{year}{2017}.
\newblock \bibinfo{title}{{Understanding the Pathophysiology of Intracranial
  Aneurysm: The ICAN Project}}.
\newblock \bibinfo{journal}{Neurosurgery} \bibinfo{volume}{80},
  \bibinfo{pages}{621--626}.
\bibitem[{Bourcier et~al.(2018)Bourcier, Scouarnec, Bonnaud
  et~al.}]{Bourcier2018}
\bibinfo{author}{Bourcier, R.}, \bibinfo{author}{Scouarnec, S.L.},
  \bibinfo{author}{Bonnaud, S.}, et~al., \bibinfo{year}{2018}.
\newblock \bibinfo{title}{Rare coding variants in angptl6 are associated with
  familial forms of intracranial aneurysm}.
\newblock \bibinfo{journal}{American Journal of Human Genetics}
  \bibinfo{volume}{102}, \bibinfo{pages}{133--141}.
\bibitem[{Chater et~al.(2021)Chater, Lauzeral, Nouri, {El Merabet} and
  Autrusseau}]{EMBC2021}
\bibinfo{author}{Chater, S.}, \bibinfo{author}{Lauzeral, N.},
  \bibinfo{author}{Nouri, A.}, \bibinfo{author}{{El Merabet}, Y.},
  \bibinfo{author}{Autrusseau, F.}, \bibinfo{year}{2021}.
\newblock \bibinfo{title}{Learning from mouse {CT}-scan brain images to detect
  {MRA-TOF} human vasculatures}, in: \bibinfo{booktitle}{43rd IEEE EMBC}, pp.
  \bibinfo{pages}{2830--2834}.
\bibitem[{Chen et~al.(2020)Chen, Wei, Lei, Liqin, Yuxin, Yakang and
  Daoying}]{chen2020}
\bibinfo{author}{Chen, G.}, \bibinfo{author}{Wei, X.}, \bibinfo{author}{Lei,
  H.}, \bibinfo{author}{Liqin, Y.}, \bibinfo{author}{Yuxin, L.},
  \bibinfo{author}{Yakang, D.}, \bibinfo{author}{Daoying, G.},
  \bibinfo{year}{2020}.
\newblock \bibinfo{title}{Automated computer-assisted detection system for
  cerebral aneurysms in time-of-flight magnetic resonance angiography using
  fully convolutional network}.
\newblock \bibinfo{journal}{BioMedical Engineering OnLine}
  \bibinfo{volume}{19}, \bibinfo{pages}{1--10}.
\bibitem[{Di~Noto et~al.(2023)Di~Noto, Marie, Tourbier, Alem{\'a}n-G{\'o}mez,
  Esteban, Saliou, Cuadra, Hagmann and Richiardi}]{di2023}
\bibinfo{author}{Di~Noto, T.}, \bibinfo{author}{Marie, G.},
  \bibinfo{author}{Tourbier, S.}, \bibinfo{author}{Alem{\'a}n-G{\'o}mez, Y.},
  \bibinfo{author}{Esteban, O.}, \bibinfo{author}{Saliou, G.},
  \bibinfo{author}{Cuadra, M.B.}, \bibinfo{author}{Hagmann, P.},
  \bibinfo{author}{Richiardi, J.}, \bibinfo{year}{2023}.
\newblock \bibinfo{title}{Towards automated brain aneurysm detection in
  tof-mra: Open data, weak labels, and anatomical knowledge}.
\newblock \bibinfo{journal}{Neuroinformatics} \bibinfo{volume}{21},
  \bibinfo{pages}{21--34}.
\bibitem[{Dierckx(1982)}]{Splines82}
\bibinfo{author}{Dierckx, P.}, \bibinfo{year}{1982}.
\newblock \bibinfo{title}{Algorithms for smoothing data with periodic and
  parametric splines}.
\newblock \bibinfo{journal}{Computer Graphics and Image Processing}
  \bibinfo{volume}{20}, \bibinfo{pages}{171--184}.
\bibitem[{Dierckx(1993)}]{Splines93}
\bibinfo{author}{Dierckx, P.}, \bibinfo{year}{1993}.
\newblock \bibinfo{title}{Curve and surface fitting with splines}.
\newblock \bibinfo{journal}{Monographs on Numerical Analysis}
  \bibinfo{volume}{63}, \bibinfo{pages}{427--428}.
\newblock \DOIprefix\doi{https://doi.org/10.2307/2153590}.
\bibitem[{Faron et~al.(2020)Faron, Sichtermann, Teichert, Luetkens, Keulers,
  Nikoubashman, Freiherr, Mpotsaris and Wiesmann}]{faron2020}
\bibinfo{author}{Faron, A.}, \bibinfo{author}{Sichtermann, T.},
  \bibinfo{author}{Teichert, N.}, \bibinfo{author}{Luetkens, J.A.},
  \bibinfo{author}{Keulers, A.}, \bibinfo{author}{Nikoubashman, O.},
  \bibinfo{author}{Freiherr, J.}, \bibinfo{author}{Mpotsaris, A.},
  \bibinfo{author}{Wiesmann, M.}, \bibinfo{year}{2020}.
\newblock \bibinfo{title}{Performance of a deep-learning neural network to
  detect intracranial aneurysms from 3d tof-mra compared to human readers}.
\newblock \bibinfo{journal}{Clinical neuroradiology} \bibinfo{volume}{30},
  \bibinfo{pages}{591--598}.
\bibitem[{Ham et~al.(2023)Ham, Seo, Yun, Bae, Kim, Sunwoo, Yoo, Jung, Kim and
  Kim}]{ham2023}
\bibinfo{author}{Ham, S.}, \bibinfo{author}{Seo, J.}, \bibinfo{author}{Yun,
  J.}, \bibinfo{author}{Bae, Y.J.}, \bibinfo{author}{Kim, T.},
  \bibinfo{author}{Sunwoo, L.}, \bibinfo{author}{Yoo, S.},
  \bibinfo{author}{Jung, S.C.}, \bibinfo{author}{Kim, J.W.},
  \bibinfo{author}{Kim, N.}, \bibinfo{year}{2023}.
\newblock \bibinfo{title}{Automated detection of intracranial aneurysms using
  skeleton-based 3d patches, semantic segmentation, and auxiliary
  classification for overcoming data imbalance in brain tof-mra}.
\newblock \bibinfo{journal}{Scientific Reports} \bibinfo{volume}{13},
  \bibinfo{pages}{12018}.
\bibitem[{Hamarneh and Jassi(2010)}]{Hamarneh2010}
\bibinfo{author}{Hamarneh, G.}, \bibinfo{author}{Jassi, P.},
  \bibinfo{year}{2010}.
\newblock \bibinfo{title}{Vascusynth: Simulating vascular trees for generating
  volumetric image data with ground truth segmentation and tree analysis}.
\newblock \bibinfo{journal}{Computerized Medical Imaging and Graphics}
  \bibinfo{volume}{34}, \bibinfo{pages}{605--616}.
\newblock \DOIprefix\doi{10.1016/j.compmedimag.2010.06.002}.
\bibitem[{Hanaoka et~al.(2019)Hanaoka, Nomura, Takenaga, Murata, Nakao, Miki,
  Yoshikawa, Hayashi, Abe and Shimizu}]{hanaoka2019}
\bibinfo{author}{Hanaoka, S.}, \bibinfo{author}{Nomura, Y.},
  \bibinfo{author}{Takenaga, T.}, \bibinfo{author}{Murata, M.},
  \bibinfo{author}{Nakao, T.}, \bibinfo{author}{Miki, S.},
  \bibinfo{author}{Yoshikawa, T.}, \bibinfo{author}{Hayashi, N.},
  \bibinfo{author}{Abe, O.}, \bibinfo{author}{Shimizu, A.},
  \bibinfo{year}{2019}.
\newblock \bibinfo{title}{Hotpig: a novel graph-based 3-d image feature set and
  its applications to computer-assisted detection of cerebral aneurysms and
  lung nodules}.
\newblock \bibinfo{journal}{International journal of computer assisted
  radiology and surgery} \bibinfo{volume}{14}, \bibinfo{pages}{2095--2107}.
\bibitem[{Haralick(1979)}]{Haralick1979}
\bibinfo{author}{Haralick, R.M.}, \bibinfo{year}{1979}.
\newblock \bibinfo{title}{Statistical and structural approaches to texture}.
\newblock \bibinfo{journal}{Proceedings of the IEEE} \bibinfo{volume}{67},
  \bibinfo{pages}{786--804}.
\bibitem[{Ivantsits et~al.(2022)Ivantsits, Goubergrits, Kuhnigk, Huellebrand,
  Bruening, Kossen, Pfahringer, Schaller, Spuler, Kuehne, Jia, Li, Shit, Menze,
  Su, Ma, Nie, Jain, Liu, Lin and Hennemuth}]{IVANTSITS2022102333}
\bibinfo{author}{Ivantsits, M.}, \bibinfo{author}{Goubergrits, L.},
  \bibinfo{author}{Kuhnigk, J.M.}, \bibinfo{author}{Huellebrand, M.},
  \bibinfo{author}{Bruening, J.}, \bibinfo{author}{Kossen, T.},
  \bibinfo{author}{Pfahringer, B.}, \bibinfo{author}{Schaller, J.},
  \bibinfo{author}{Spuler, A.}, \bibinfo{author}{Kuehne, T.},
  \bibinfo{author}{Jia, Y.}, \bibinfo{author}{Li, X.}, \bibinfo{author}{Shit,
  S.}, \bibinfo{author}{Menze, B.}, \bibinfo{author}{Su, Z.},
  \bibinfo{author}{Ma, J.}, \bibinfo{author}{Nie, Z.}, \bibinfo{author}{Jain,
  K.}, \bibinfo{author}{Liu, Y.}, \bibinfo{author}{Lin, Y.},
  \bibinfo{author}{Hennemuth, A.}, \bibinfo{year}{2022}.
\newblock \bibinfo{title}{Detection and analysis of cerebral aneurysms based on
  x-ray rotational angiography - the cada 2020 challenge}.
\newblock \bibinfo{journal}{Medical Image Analysis} \bibinfo{volume}{77},
  \bibinfo{pages}{102333}.
\bibitem[{Joo et~al.(2020)Joo, Ahn, Yoon, Bae, Sohn, Lee, Bae, Park, Choi and
  Lee}]{joo2020}
\bibinfo{author}{Joo, B.}, \bibinfo{author}{Ahn, S.S.}, \bibinfo{author}{Yoon,
  P.H.}, \bibinfo{author}{Bae, S.}, \bibinfo{author}{Sohn, B.},
  \bibinfo{author}{Lee, Y.E.}, \bibinfo{author}{Bae, J.H.},
  \bibinfo{author}{Park, M.S.}, \bibinfo{author}{Choi, H.S.},
  \bibinfo{author}{Lee, S.K.}, \bibinfo{year}{2020}.
\newblock \bibinfo{title}{A deep learning algorithm may automate intracranial
  aneurysm detection on mr angiography with high diagnostic performance}.
\newblock \bibinfo{journal}{European Radiology} \bibinfo{volume}{30},
  \bibinfo{pages}{5785--5793}.
\bibitem[{Karch et~al.(1999)Karch, Neumann, Neumann and Schreiner}]{Karch1999}
\bibinfo{author}{Karch, R.}, \bibinfo{author}{Neumann, F.},
  \bibinfo{author}{Neumann, M.}, \bibinfo{author}{Schreiner, W.},
  \bibinfo{year}{1999}.
\newblock \bibinfo{title}{A three-dimensional model for arterial tree
  representation, generated by constrained constructive optimization}.
\newblock \bibinfo{journal}{Computers in Biology and Medicine}
  \bibinfo{volume}{29}, \bibinfo{pages}{19--38}.
\newblock \URLprefix
  \url{https://www.sciencedirect.com/science/article/pii/S0010482598000456},
  \DOIprefix\doi{https://doi.org/10.1016/S0010-4825(98)00045-6}.
\bibitem[{Kretowski et~al.(2003)Kretowski, Rolland, Bezy-Wendling and
  Coatrieux}]{Coatrieux2003}
\bibinfo{author}{Kretowski, M.}, \bibinfo{author}{Rolland, Y.},
  \bibinfo{author}{Bezy-Wendling, J.}, \bibinfo{author}{Coatrieux, J.L.},
  \bibinfo{year}{2003}.
\newblock \bibinfo{title}{Physiologically based modeling of 3-d vascular
  networks and ct scan angiography}.
\newblock \bibinfo{journal}{IEEE Transactions on Medical Imaging}
  \bibinfo{volume}{22}, \bibinfo{pages}{248--257}.
\newblock \DOIprefix\doi{10.1109/TMI.2002.808357}.
\bibitem[{L'Allinec et~al.(2020)L'Allinec, Chatel, Karakachoff, Bourcereau,
  Lamoureux, Gaignard, Autrusseau, Jouan, Vion, Loirand, Desal, Naggara, Redon,
  Edjlali and Bourcier}]{Neurosurg2020}
\bibinfo{author}{L'Allinec, V.}, \bibinfo{author}{Chatel, S.},
  \bibinfo{author}{Karakachoff, M.}, \bibinfo{author}{Bourcereau, E.},
  \bibinfo{author}{Lamoureux, Z.}, \bibinfo{author}{Gaignard, A.},
  \bibinfo{author}{Autrusseau, F.}, \bibinfo{author}{Jouan, S.},
  \bibinfo{author}{Vion, A.C.}, \bibinfo{author}{Loirand, G.},
  \bibinfo{author}{Desal, H.}, \bibinfo{author}{Naggara, O.},
  \bibinfo{author}{Redon, R.}, \bibinfo{author}{Edjlali, M.},
  \bibinfo{author}{Bourcier, R.}, \bibinfo{year}{2020}.
\newblock \bibinfo{title}{Unruptured cerebral aneurysm, prediction of evolution
  : The {UCAN} project}.
\newblock \bibinfo{journal}{Neurosurgery, Oxford University Press} .
\bibitem[{Lauric et~al.(2010)Lauric, Miller, Frisken and Malek}]{LAURIC2010149}
\bibinfo{author}{Lauric, A.}, \bibinfo{author}{Miller, E.},
  \bibinfo{author}{Frisken, S.}, \bibinfo{author}{Malek, A.M.},
  \bibinfo{year}{2010}.
\newblock \bibinfo{title}{Automated detection of intracranial aneurysms based
  on parent vessel 3d analysis}.
\newblock \bibinfo{journal}{Medical Image Analysis} \bibinfo{volume}{14},
  \bibinfo{pages}{149--159}.
\newblock \URLprefix
  \url{https://www.sciencedirect.com/science/article/pii/S1361841509001212},
  \DOIprefix\doi{https://doi.org/10.1016/j.media.2009.10.005}.
\bibitem[{Liao et~al.(2001)Liao, Chen and Chung}]{Liao2001}
\bibinfo{author}{Liao, P.S.}, \bibinfo{author}{Chen, T.S.},
  \bibinfo{author}{Chung, P.C.}, \bibinfo{year}{2001}.
\newblock \bibinfo{title}{A fast algorithm for multilevel thresholding}.
\newblock \bibinfo{journal}{Journal of Information Science and Engineering}
  \bibinfo{volume}{17(5)}, \bibinfo{pages}{713--727}.
\bibitem[{Nader et~al.(2023)Nader, Bourcier and Autrusseau}]{NADER2023102919}
\bibinfo{author}{Nader, R.}, \bibinfo{author}{Bourcier, R.},
  \bibinfo{author}{Autrusseau, F.}, \bibinfo{year}{2023}.
\newblock \bibinfo{title}{Using deep learning for an automatic detection and
  classification of the vascular bifurcations along the circle of willis}.
\newblock \bibinfo{journal}{Medical Image Analysis} ,
  \bibinfo{pages}{102919}\URLprefix
  \url{https://www.sciencedirect.com/science/article/pii/S1361841523001792},
  \DOIprefix\doi{https://doi.org/10.1016/j.media.2023.102919}.
\bibitem[{Nakao et~al.(2018)Nakao, Hanaoka, Nomura, Sato, Nemoto, Miki, Maeda,
  Yoshikawa, Hayashi and Abe}]{nakao2018}
\bibinfo{author}{Nakao, T.}, \bibinfo{author}{Hanaoka, S.},
  \bibinfo{author}{Nomura, Y.}, \bibinfo{author}{Sato, I.},
  \bibinfo{author}{Nemoto, M.}, \bibinfo{author}{Miki, S.},
  \bibinfo{author}{Maeda, E.}, \bibinfo{author}{Yoshikawa, T.},
  \bibinfo{author}{Hayashi, N.}, \bibinfo{author}{Abe, O.},
  \bibinfo{year}{2018}.
\newblock \bibinfo{title}{Deep neural network-based computer-assisted detection
  of cerebral aneurysms in mr angiography}.
\newblock \bibinfo{journal}{Journal of Magnetic Resonance Imaging}
  \bibinfo{volume}{47}, \bibinfo{pages}{948--953}.
\bibitem[{Nemoto et~al.(2017)Nemoto, Hayashi, Hanaoka, Nomura, Miki and
  Yoshikawa}]{nemoto2017feasibility}
\bibinfo{author}{Nemoto, M.}, \bibinfo{author}{Hayashi, N.},
  \bibinfo{author}{Hanaoka, S.}, \bibinfo{author}{Nomura, Y.},
  \bibinfo{author}{Miki, S.}, \bibinfo{author}{Yoshikawa, T.},
  \bibinfo{year}{2017}.
\newblock \bibinfo{title}{Feasibility study of a generalized framework for
  developing computer-aided detection systems—a new paradigm}.
\newblock \bibinfo{journal}{Journal of digital imaging} \bibinfo{volume}{30},
  \bibinfo{pages}{629--639}.
\bibitem[{Nouri et~al.(2018)Nouri, Autrusseau and
  Bourcier}]{Bifurc_patent_2018}
\bibinfo{author}{Nouri, A.}, \bibinfo{author}{Autrusseau, F.},
  \bibinfo{author}{Bourcier, R.}, \bibinfo{year}{2018}.
\newblock \bibinfo{title}{Method for locating and characterizing bifurcations
  of a cerebral vascular tree, associated methods and devices}.
\newblock \URLprefix
  \url{https://patentscope.wipo.int/search/en/detail.jsf?docId=WO2020115162}.
\bibitem[{Nouri et~al.(2020)Nouri, Autrusseau, Bourcier, Gaignard, {L'Allinec},
  Menguy, Veziers, Desal, Loirand and Redon}]{Nouri2020}
\bibinfo{author}{Nouri, A.}, \bibinfo{author}{Autrusseau, F.},
  \bibinfo{author}{Bourcier, R.}, \bibinfo{author}{Gaignard, A.},
  \bibinfo{author}{{L'Allinec}, V.}, \bibinfo{author}{Menguy, C.},
  \bibinfo{author}{Veziers, J.}, \bibinfo{author}{Desal, H.},
  \bibinfo{author}{Loirand, G.}, \bibinfo{author}{Redon, R.},
  \bibinfo{year}{2020}.
\newblock \bibinfo{title}{Characterization of {3D} bifurcations in micro-scan
  and {MRA-TOF} images of cerebral vasculature for prediction of intra-cranial
  aneurysms}.
\newblock \bibinfo{journal}{Elsevier Computerized Medical Imaging and Graphics}
  \bibinfo{volume}{84C}.
\bibitem[{Park et~al.(2019)Park, Chute, Rajpurkar, Lou, Ball, Shpanskaya,
  Jabarkheel, Kim, McKenna, Tseng, Ni, Wishah, Wittber, Hong, Wilson, Halabi,
  Basu, Patel, Lungren, Ng and Yeom}]{Park2019}
\bibinfo{author}{Park, A.}, \bibinfo{author}{Chute, C.},
  \bibinfo{author}{Rajpurkar, P.}, \bibinfo{author}{Lou, J.},
  \bibinfo{author}{Ball, R.L.}, \bibinfo{author}{Shpanskaya, K.S.},
  \bibinfo{author}{Jabarkheel, R.}, \bibinfo{author}{Kim, L.H.},
  \bibinfo{author}{McKenna, E.}, \bibinfo{author}{Tseng, J.},
  \bibinfo{author}{Ni, J.C.}, \bibinfo{author}{Wishah, F.},
  \bibinfo{author}{Wittber, F.}, \bibinfo{author}{Hong, D.S.},
  \bibinfo{author}{Wilson, T.J.}, \bibinfo{author}{Halabi, S.S.},
  \bibinfo{author}{Basu, S.}, \bibinfo{author}{Patel, B.N.},
  \bibinfo{author}{Lungren, M.P.}, \bibinfo{author}{Ng, A.},
  \bibinfo{author}{Yeom, K.W.}, \bibinfo{year}{2019}.
\newblock \bibinfo{title}{Deep learning–assisted diagnosis of cerebral
  aneurysms using the headxnet model}.
\newblock \bibinfo{journal}{JAMA Network Open} \bibinfo{volume}{2}.
\newblock \URLprefix \url{https://api.semanticscholar.org/CorpusID:174811733}.
\bibitem[{Robben et~al.(2016)Robben, T{\"{u}}retken, Sunaert, Thijs, Wilms,
  Fua, Maes and Suetens}]{Robben2016}
\bibinfo{author}{Robben, D.}, \bibinfo{author}{T{\"{u}}retken, E.},
  \bibinfo{author}{Sunaert, S.}, \bibinfo{author}{Thijs, V.},
  \bibinfo{author}{Wilms, G.}, \bibinfo{author}{Fua, P.},
  \bibinfo{author}{Maes, F.}, \bibinfo{author}{Suetens, P.},
  \bibinfo{year}{2016}.
\newblock \bibinfo{title}{{Simultaneous segmentation and anatomical labeling of
  the cerebral vasculature}}.
\newblock \bibinfo{journal}{Medical Image Analysis} \bibinfo{volume}{32},
  \bibinfo{pages}{201--215}.
\bibitem[{Ronneberger et~al.(2015)Ronneberger, Fischer and
  Brox}]{Ronneberger2015}
\bibinfo{author}{Ronneberger, O.}, \bibinfo{author}{Fischer, P.},
  \bibinfo{author}{Brox, T.}, \bibinfo{year}{2015}.
\newblock \bibinfo{title}{{U-Net}: Convolutional networks for biomedical image
  segmentation}, in: \bibinfo{booktitle}{Medical Image Computing and
  Computer-Assisted Intervention (MICCAI)}, \bibinfo{publisher}{Springer}. pp.
  \bibinfo{pages}{234--241}.
\bibitem[{Sailer et~al.(2014)Sailer, Wagemans, Nelemans, de~Graaf and van
  Zwam}]{sailer2014}
\bibinfo{author}{Sailer, A.M.}, \bibinfo{author}{Wagemans, B.A.},
  \bibinfo{author}{Nelemans, P.J.}, \bibinfo{author}{de~Graaf, R.},
  \bibinfo{author}{van Zwam, W.H.}, \bibinfo{year}{2014}.
\newblock \bibinfo{title}{Diagnosing intracranial aneurysms with mr
  angiography: systematic review and meta-analysis}.
\newblock \bibinfo{journal}{Stroke} \bibinfo{volume}{45},
  \bibinfo{pages}{119--126}.
\bibitem[{Shi et~al.(2020)Shi, Miao, Schoepf, Savage, Dargis, Pan, Chai, Li,
  Xia, Zhang et~al.}]{natu}
\bibinfo{author}{Shi, Z.}, \bibinfo{author}{Miao, C.},
  \bibinfo{author}{Schoepf, U.J.}, \bibinfo{author}{Savage, R.H.},
  \bibinfo{author}{Dargis, D.M.}, \bibinfo{author}{Pan, C.},
  \bibinfo{author}{Chai, X.}, \bibinfo{author}{Li, X.L.}, \bibinfo{author}{Xia,
  S.}, \bibinfo{author}{Zhang, X.}, et~al., \bibinfo{year}{2020}.
\newblock \bibinfo{title}{A clinically applicable deep-learning model for
  detecting intracranial aneurysm in computed tomography angiography images}.
\newblock \bibinfo{journal}{Nature communications} \bibinfo{volume}{11},
  \bibinfo{pages}{6090}.
\bibitem[{Sichtermann et~al.(2019)Sichtermann, Faron, Sijben, Teichert,
  Freiherr and Wiesmann}]{Sichtermann2019}
\bibinfo{author}{Sichtermann, T.}, \bibinfo{author}{Faron, A.},
  \bibinfo{author}{Sijben, R.}, \bibinfo{author}{Teichert, N.},
  \bibinfo{author}{Freiherr, J.}, \bibinfo{author}{Wiesmann, M.},
  \bibinfo{year}{2019}.
\newblock \bibinfo{title}{Deep learning-based detection of intracranial
  aneurysms in {3D} {TOF-MRA}}.
\newblock \bibinfo{journal}{American Journal of Neuroradiology}
  \bibinfo{volume}{40}, \bibinfo{pages}{25--32}.
\bibitem[{Simard et~al.(2003)Simard, Steinkraus and Platt}]{Simard2003}
\bibinfo{author}{Simard, P.}, \bibinfo{author}{Steinkraus, D.},
  \bibinfo{author}{Platt, J.}, \bibinfo{year}{2003}.
\newblock \bibinfo{title}{Best practices for convolutional neural networks
  applied to visual document analysis}, in: \bibinfo{booktitle}{Seventh
  International Conference on Document Analysis and Recognition, 2003.
  Proceedings.}, pp. \bibinfo{pages}{958--963}.
\newblock \DOIprefix\doi{10.1109/ICDAR.2003.1227801}.
\bibitem[{Stember et~al.(2019)Stember, Chang, Stember, Liu, Grinband, Filippi,
  Meyers and Jambawalikar}]{stember2019}
\bibinfo{author}{Stember, J.N.}, \bibinfo{author}{Chang, P.},
  \bibinfo{author}{Stember, D.M.}, \bibinfo{author}{Liu, M.},
  \bibinfo{author}{Grinband, J.}, \bibinfo{author}{Filippi, C.G.},
  \bibinfo{author}{Meyers, P.}, \bibinfo{author}{Jambawalikar, S.},
  \bibinfo{year}{2019}.
\newblock \bibinfo{title}{Convolutional neural networks for the detection and
  measurement of cerebral aneurysms on magnetic resonance angiography}.
\newblock \bibinfo{journal}{Journal of digital imaging} \bibinfo{volume}{32},
  \bibinfo{pages}{808--815}.
\bibitem[{Summerlin et~al.(2022)Summerlin, Willis, Boggs, Johnson and
  Porter}]{Summerlin2022}
\bibinfo{author}{Summerlin, D.}, \bibinfo{author}{Willis, J.},
  \bibinfo{author}{Boggs, R.}, \bibinfo{author}{Johnson, L.},
  \bibinfo{author}{Porter, K.}, \bibinfo{year}{2022}.
\newblock \bibinfo{title}{Radiation dose reduction opportunities in vascular
  imaging}.
\newblock \bibinfo{journal}{Tomography} \bibinfo{volume}{8},
  \bibinfo{pages}{2618--2638}.
\newblock \DOIprefix\doi{http://dx.doi.org/10.3390/tomography8050219}.
\bibitem[{Szczerba and Sz{\'e}kely(2002)}]{Szekely2002}
\bibinfo{author}{Szczerba, D.}, \bibinfo{author}{Sz{\'e}kely, G.},
  \bibinfo{year}{2002}.
\newblock \bibinfo{title}{Macroscopic modeling of vascular systems}, in:
  \bibinfo{editor}{Dohi, T.}, \bibinfo{editor}{Kikinis, R.} (Eds.),
  \bibinfo{booktitle}{Medical Image Computing and Computer-Assisted
  Intervention --- MICCAI 2002}, \bibinfo{publisher}{Springer Berlin
  Heidelberg}, \bibinfo{address}{Berlin, Heidelberg}. pp.
  \bibinfo{pages}{284--292}.
\bibitem[{Taghanaki et~al.(2019)Taghanaki, Zheng, Zhou, Georgescu, Sharma, Xu,
  Comaniciu and Hamarneh}]{Taghanaki2019}
\bibinfo{author}{Taghanaki, S.}, \bibinfo{author}{Zheng, Y.},
  \bibinfo{author}{Zhou, K.}, \bibinfo{author}{Georgescu, B.},
  \bibinfo{author}{Sharma, P.}, \bibinfo{author}{Xu, D.},
  \bibinfo{author}{Comaniciu, D.}, \bibinfo{author}{Hamarneh, G.},
  \bibinfo{year}{2019}.
\newblock \bibinfo{title}{Combo loss: Handling input and output imbalance in
  multi-organ segmentation}.
\newblock \bibinfo{journal}{Comput Med Imaging Graph} \bibinfo{volume}{75},
  \bibinfo{pages}{24--33}.
\newblock \DOIprefix\doi{10.1016/j.compmedimag.2019.04.005}.
\bibitem[{Timmins et~al.(2021)Timmins, {van der Schaaf}, Bennink
  et~al.}]{Timmins2021}
\bibinfo{author}{Timmins, K.}, \bibinfo{author}{{van der Schaaf}, I.},
  \bibinfo{author}{Bennink, E.}, et~al., \bibinfo{year}{2021}.
\newblock \bibinfo{title}{Comparing methods of detecting and segmenting
  unruptured intracranial aneurysms on {TOF-MRAS}: The {ADAM} challenge}.
\newblock \bibinfo{journal}{NeuroImage} \bibinfo{volume}{238},
  \bibinfo{pages}{118216}.
\bibitem[{Ueda et~al.(2019)Ueda, Yamamoto, Nishimori et~al.}]{Ueda2019}
\bibinfo{author}{Ueda, D.}, \bibinfo{author}{Yamamoto, A.},
  \bibinfo{author}{Nishimori, M.}, et~al., \bibinfo{year}{2019}.
\newblock \bibinfo{title}{Deep learning for {MR} angiography: Automated
  detection of cerebral aneurysms}.
\newblock \bibinfo{journal}{Radiology} \bibinfo{volume}{290},
  \bibinfo{pages}{187--194}.
\newblock \DOIprefix\doi{10.1148/radiol.2018180901}. \bibinfo{note}{pMID:
  30351253}.
\bibitem[{Wang and Summers(2012)}]{wang2012machine}
\bibinfo{author}{Wang, S.}, \bibinfo{author}{Summers, R.M.},
  \bibinfo{year}{2012}.
\newblock \bibinfo{title}{Machine learning and radiology}.
\newblock \bibinfo{journal}{Medical image analysis} \bibinfo{volume}{16},
  \bibinfo{pages}{933--951}.
\bibitem[{Yang et~al.(2021)Yang, Xie, Hu, Alwalid, Xu, Liu, Jin, Li, Tu, Liu
  et~al.}]{yang2021deep}
\bibinfo{author}{Yang, J.}, \bibinfo{author}{Xie, M.}, \bibinfo{author}{Hu,
  C.}, \bibinfo{author}{Alwalid, O.}, \bibinfo{author}{Xu, Y.},
  \bibinfo{author}{Liu, J.}, \bibinfo{author}{Jin, T.}, \bibinfo{author}{Li,
  C.}, \bibinfo{author}{Tu, D.}, \bibinfo{author}{Liu, X.}, et~al.,
  \bibinfo{year}{2021}.
\newblock \bibinfo{title}{Deep learning for detecting cerebral aneurysms with
  ct angiography}.
\newblock \bibinfo{journal}{Radiology} \bibinfo{volume}{298},
  \bibinfo{pages}{155--163}.
\bibitem[{Yang et~al.(2011)Yang, Blezek, Cheng, Ryan, Kallmes and
  Erickson}]{yang2011computer}
\bibinfo{author}{Yang, X.}, \bibinfo{author}{Blezek, D.J.},
  \bibinfo{author}{Cheng, L.T.}, \bibinfo{author}{Ryan, W.J.},
  \bibinfo{author}{Kallmes, D.F.}, \bibinfo{author}{Erickson, B.J.},
  \bibinfo{year}{2011}.
\newblock \bibinfo{title}{Computer-aided detection of intracranial aneurysms in
  mr angiography}.
\newblock \bibinfo{journal}{Journal of digital imaging} \bibinfo{volume}{24},
  \bibinfo{pages}{86--95}.
\bibitem[{Zeng et~al.(2020)Zeng, Liu, Xiao, Li, Jiang, Feng and Guo}]{zeng}
\bibinfo{author}{Zeng, Y.}, \bibinfo{author}{Liu, X.}, \bibinfo{author}{Xiao,
  N.}, \bibinfo{author}{Li, Y.}, \bibinfo{author}{Jiang, Y.},
  \bibinfo{author}{Feng, J.}, \bibinfo{author}{Guo, S.}, \bibinfo{year}{2020}.
\newblock \bibinfo{title}{Automatic diagnosis based on spatial information
  fusion feature for intracranial aneurysm}.
\newblock \bibinfo{journal}{IEEE Transactions on Medical Imaging}
  \bibinfo{volume}{39}, \bibinfo{pages}{1448--1458}.
\newblock \DOIprefix\doi{10.1109/TMI.2019.2951439}.
\bibitem[{Zhou et~al.(2018)Zhou, Dion and Rouleau}]{Zhou2018}
\bibinfo{author}{Zhou, S.}, \bibinfo{author}{Dion, P.A.},
  \bibinfo{author}{Rouleau, G.A.}, \bibinfo{year}{2018}.
\newblock \bibinfo{title}{Genetics of intracranial aneurysms}.
\newblock \bibinfo{journal}{Stroke} \bibinfo{volume}{49},
  \bibinfo{pages}{780–787}.
\newblock \DOIprefix\doi{https://doi.org/10.1161/STROKEAHA.117.018152}.
\bibitem[{Çiçek et~al.(2016)Çiçek, Abdulkadir, Lienkamp, Brox and
  Ronneberger}]{cicek20163d}
\bibinfo{author}{Çiçek, O.}, \bibinfo{author}{Abdulkadir, A.},
  \bibinfo{author}{Lienkamp, S.S.}, \bibinfo{author}{Brox, T.},
  \bibinfo{author}{Ronneberger, O.}, \bibinfo{year}{2016}.
\newblock \bibinfo{title}{{3D U-Net}: Learning dense volumetric segmentation
  from sparse annotation}, in: \bibinfo{booktitle}{Medical Image Computing and
  Computer-Assisted Intervention (MICCAI)}, \bibinfo{publisher}{Springer
  International Publishing}. pp. \bibinfo{pages}{424--432}.
\newblock \href{http://arxiv.org/abs/1606.06650}{\tt arXiv:1606.06650}.

\end{thebibliography}

\end{document}